\documentclass[12pt,letterpaper]{article}
\usepackage{amsfonts}
\usepackage{amssymb}
\usepackage{amsmath}
\usepackage{amsthm}
\usepackage[round,authoryear]{natbib}
\usepackage[margin=2.2cm]{geometry}
\usepackage{booktabs}
\usepackage{dcolumn}
\newcolumntype{d}[1]{D{.}{.}{#1}} 
\usepackage{graphicx,epstopdf}
\usepackage[T1]{fontenc}
\usepackage{lmodern}
\usepackage{enumitem}
\setlist[enumerate,1]{itemsep=1pt, topsep=1pt, partopsep=0pt, parsep=1pt}
\setlist[enumerate,2]{nosep}
\setlist[itemize,1]{itemsep=1pt, topsep=1pt, partopsep=0pt, parsep=1pt}
\setlist[itemize,2]{nosep}
\usepackage{subfig}
\usepackage{bm}
\usepackage[colorlinks,linkcolor=red,citecolor=blue,pdftex,breaklinks]{hyperref}
\hypersetup{bookmarksopen=true}
\usepackage[nameinlink]{cleveref}
\usepackage{setspace}

\makeatletter
\AddToHook{env/figure/begin}{\def\ltx@label#1{\cref@label{#1}}}
\AddToHook{env/table/begin}{\def\ltx@label#1{\cref@label{#1}}}
\makeatother

\makeatletter
\AddToHook{cmd/appendix/before}{\def\cref@section@alias{appendix}}
\makeatother

\theoremstyle{plain}
\newtheorem{theorem}{Theorem}

\newtheorem{lemma}{Lemma}[section]
\theoremstyle{definition}
\newtheorem{assumption}{Assumption}

\Crefname{assumptionx}{Assumption}{Assumptions} 
\Crefname{assumption}{Assumption}{Assumptions} 
\Crefname{examplex}{Example}{Examples} 
\Crefname{remarkx}{Remark}{Remarks} 
\Crefname{appendix}{Appendix}{Appendices} 

\makeatletter
\renewcommand*{\eqref}[1]{\hyperref[{#1}]{\textup{\tagform@{\ref*{#1}}}}}
\makeatother

\makeatletter
\DeclareRobustCommand\citepos
  {\begingroup\def\NAT@nmfmt##1{{\NAT@up##1's}}%
   \NAT@swafalse\let\NAT@ctype\z@\NAT@partrue
   \@ifstar{\NAT@fulltrue\NAT@citetp}{\NAT@fullfalse\NAT@citetp}}
\makeatother

\expandafter
\def \expandafter \normalsize \expandafter{\normalsize \setlength \abovedisplayskip{8pt plus 2pt minus 7pt}}
\expandafter
\def \expandafter \normalsize \expandafter{\normalsize \setlength \abovedisplayshortskip{0pt plus 2pt}}
\expandafter
\def \expandafter \normalsize \expandafter{\normalsize \setlength \belowdisplayskip{8pt plus 2pt minus 7pt}}
\expandafter
\def \expandafter \normalsize \expandafter{\normalsize \setlength \belowdisplayshortskip{3pt plus 1pt minus 2pt}}

\def\bia{{\bm{a}}}

\def\bbeta{{\bm\beta}}
\def\bgamma{{\bm\gamma}}

\def\biX{{\bm{X}}}
\def\biZ{{\bm{Z}}}
\def\biW{{\bm{W}}}
\def\biM{{\bm{M}}}

\def\biD{{\bm{D}}}

\def\biR{{\bm{R}}}
\def\biS{{\bm{S}}}

\def\bir{{\bm{r}}}
\def\bis{{\bm{s}}}
\def\biy{{\bm{y}}}

\def\biu{{\bm{u}}}

\def\bSigma{{\bm{\Sigma}}}
\def\bLambda{{\bm{\Lambda}}}
\def\biU{{\bm{U}}}
\def\biV{{\bm{V}}}

\def\bdelta{{\bm{\delta}}}
\def\bmu{{\bm{\mu}}}

\def\bxi{{\bm{\xi}}}
\def\E{{\rm E}}

\def\N{{\rm N}}
\def\dto{\overset{d}  \longrightarrow}
\def\Pto{\overset{P}  \longrightarrow}

\def\tk{\kern 0.08333em}
\def\tn{\kern -0.08333em}
\def\tkk{\kern 0.04167em}
\def\bzero{{\bm{0}}}
\def\bfI{{\bf I}}

\def\th#1{$#1^{\tk{\rm th}}$}

\DeclareMathOperator{\Var}{Var}

\newcommand{\FO}[1]{}

\usepackage[small,compact]{titlesec}

\begin{document}

\title{Jackknife Inference with Two\tkk-\tn Way Clustering\thanks{We
are grateful to conference and seminar participants at the Canadian 
Econometric Study Group, University of Graz, UCLA, the Aarhus Workshop 
in Econometrics, the 2025 Canadian Stata Meeting, and the Workshop 
for Lynda Khalaf. MacKinnon and Webb thank the Social Sciences and 
Humanities Research Council of Canada (SSHRC grant 435-2021-0396), 
and Nielsen thanks the Danish National Research Foundation (DNRF Chair 
grant number DNRF154) for financial support. MacKinnon and Nielsen are 
also grateful for support from the Aarhus Center for Econometrics 
(ACE) funded by the Danish National Research Foundation grant number 
DNRF186. Code and data files may be found at
\url{http://qed.econ.queensu.ca/pub/faculty/mackinnon/twowayjack/} }}

\author{James G. MacKinnon\thanks{Corresponding author.}\\
{\small Dept.\ of Economics, Queen's University}\\ 
{\small Aarhus Center for Econometrics, Aarhus University}\\ 
\texttt{mackinno@queensu.ca} \and
Morten \O rregaard Nielsen\\{\small Aarhus Center for Econometrics, 
Aarhus University}\\ \texttt{mon@econ.au.dk} \and Matthew D. Webb\\
{\small Dept.\ of Economics, Carleton University}\\ 
\texttt{matt.webb@carleton.ca}}

\maketitle

\begin{abstract}
For linear regression models with cross\tkk-section or panel data, it
is natural to assume that the disturbances are clustered in two
dimensions. However, the finite\tkk-sample properties of two\tkk-way
cluster-robust tests and confidence intervals are often poor. We
discuss several ways to improve inference with two\tkk-way clustering.
Two of these are existing methods for avoiding, or at least
ameliorating, the problem of undefined standard errors when a
cluster-robust variance matrix estimator (CRVE) is not positive
definite. One is a new method that always avoids the problem. More
importantly, we propose a family of new two\tkk-way CRVEs based on the
cluster jackknife and prove that they yield valid inferences
asymptotically. Simulations for models with two\tkk-way fixed effects
suggest that, in many cases, the cluster-jackknife CRVE combined with
our new method yields surprisingly accurate inferences. We provide a
software package, \texttt{twowayjack} for \texttt{Stata}, that
implements our recommended variance estimator.

\vskip 12pt



\end{abstract}

\clearpage
\onehalfspacing

\section{Introduction}
\label{sec:intro}

The use of two\tkk-way cluster-robust variance estimators for linear
regression models was independently proposed by \citet{MH_2006},
\citet*{CGM_2011}, and \citet{Thompson_2011}. Although two\tkk-way
clustering has been widely used in empirical work, the asymptotic
theory to justify it is quite recent. See, among others,
\citet*{Davezies_2021,Davezies_2025}, \citet*{MNW_2021},
\citet{Menzel_2021}, \citet*{Chiang_2022-JBES}, \citet*{CKS_2023},
\citet*{CHS_2024}, and \citet{yap_2025}. The finite\tkk-sample
properties of statistical inference are much less well understood for
two\tkk-way clustering than for one\tkk-way clustering. For an
up-to\tkk-date discussion of the latter, with recommendations for
empirical practice, see \citet*{MNW-guide}.

The jackknife variance estimator has been around for a very long time
\citep{Tukey_1958,Efron_81,Efron-Stein}. The cluster-jackknife CRVE
(sometimes called the CV$_{\tn3}$ estimator) for linear regression
models with one\tkk-way clustering was proposed in \citet{BM_2002} and
has been available in \texttt{Stata} for many years. Nevertheless, it
has not been studied or applied much until very recently. In part,
this is because \citet{BM_2002} followed \citet{MW_1985} by computing
the CV$_{\tn3}$ estimator in a way that is efficient when all clusters
are very small but extremely inefficient when any clusters are large;
see \citet*{MNW-bootknife}. This seems to have given many
investigators the erroneous impression that CV$_{\tn3}$ is very
expensive to compute, even though the \texttt{Stata} implementation
uses a method that is reasonably efficient when the number of clusters
is not too large. An even more efficient method is discussed in
\citet*{MNW-influence} and implemented in the \texttt{Stata} package
\texttt{summclust} \citep*{MNW_summclust}.

In \Cref{sec:concepts}, we discuss the linear regression model with
two\tkk-way clustering. Two existing CRVEs are discussed, along with
their theoretical and practical deficiencies. For the CRVE that is
theoretically soundest, the chief deficiency is that it may not be
positive definite in finite samples. We discuss two ways to overcome
this problem. One is the eigen-decomposition method suggested in
\citet{CGM_2011}. The other is a new and extremely simple procedure
which can readily be implemented using existing software.

In \Cref{sec:twojack}, we show how to extend the cluster-jackknife
CRVEs discussed in \citet{MNW-bootknife} and \citet{Hansen-jack} to
two\tkk-way clustering. Two alternative approaches, based on different
jackknife constructions, have been proposed very recently; see
\citet*{CMO_2025}, which uses empirical likelihood, and
\citet{Hounyo-jack}. In \Cref{subsec:consistency}, we prove that our
two\tkk-way cluster-jackknife CRVE is consistent. Based on what is
known about the finite\tkk-sample performance of cluster-jackknife
CRVEs for one\tkk-way clustering, it seems very likely that inference
based on our CRVE will be more conservative, and usually more
reliable, than conventional inference in the two\tkk-way case as well.
Some theoretical arguments to support this conjecture are provided in
\Cref{subsec:whyjack}.

In \Cref{sec:simuls}, we use simulation experiments to study the
finite\tkk-sample performance of several procedures for inference.
Using the cluster-jackknife methods of \Cref{sec:twojack} in
combination with either of the procedures discussed in
\Cref{sec:concepts} often performs much better than existing methods
for cluster-robust inference. In \Cref{sec:empirical}, we apply
several methods to two empirical examples. The results that we obtain
are entirely in accord with the simulations in \Cref{sec:simuls}. We
conclude that, while conventional methods probably do not yield
reliable inferences for these examples, our preferred methods based on
the cluster jackknife probably do. Finally, \Cref{sec:conclusions}
concludes.

\section{Cluster-Robust Variance Estimation in Two Dimensions}
\label{sec:concepts}

Consider the linear regression model
\begin{equation}
\label{model} 
\biy = \biX\tn\bbeta + \biu ,
\end{equation}
where $\biy$ and $\biu$ are $N\times 1$ vectors of observations and
disturbances, $\biX$ is an $N\times k$ matrix of covariates, and
$\bbeta$ is a $k\times1$ parameter vector. The model is assumed to
have two dimensions of clustering, where the numbers of clusters in
the two dimensions are $G$ and $H$\tn, respectively. It is
illuminating to rewrite \eqref{model} in terms of the intersections of
the two clustering dimensions:
\begin{equation}
\label{modelgh}
\biy_{gh} = \biX_{gh}\bbeta + \biu_{gh},\quad g=1,\ldots,G,
\;\; h=1,\ldots,H.
\end{equation}
Here the vectors $\biy_{gh}$ and $\biu_{gh}$ and the matrix
$\biX_{gh}$ contain, respectively, the rows of $\biy$, $\biu$, and
$\biX$ that correspond to both the \th{g} cluster in the first
clustering dimension and the \th{h} cluster in the second clustering
dimension. Similarly, we use $\biy_g$, $\biX_g$, and $\biu_g$ to denote 
vectors that contain the rows of $\biy$, $\biX$\tn, and $\biu$ for the 
\th{g} cluster in the first dimension, and $\biy_h$, $\biX_h$, and 
$\biu_h$ to denote the corresponding rows for the \th{h} cluster in the 
second dimension. For example, the vector $\biy_g$ contains the 
subvectors $\biy_{g1}$ through~$\biy_{gH}$.

We use $N_g$ to denote the number of observations in cluster $g$ for
the first dimension, $N_h$ to denote the number of observations in
cluster $h$ for the second dimension, and $N_{gh}$ to denote the
number of observations in the intersection of cluster $g$ in the first
dimension with cluster $h$ in the second dimension. We assume that
$N_g\ge1$ and $N_h\ge1$. Thus, the number of observations in the entire 
sample is
\begin{equation*}
N=\sum_{g=1}^G N_g=\sum_{h=1}^H N_h=\sum_{g=1}^G \sum_{h=1}^H N_{gh}.
\end{equation*}
Note that some of the intersections may be empty, so that $N_{gh}$ might
well equal 0 for some values of $g$ and~$h$. The number of non-empty 
intersections is $I \le GH$.

Various score vectors play key roles in cluster-robust inference. The
score vector for the entire sample is $\bis = \biX^\top\biu$. The
score subvector for cluster $g$ in the first dimension is $\bis_g =
\biX_g^\top\biu_g$, and the score subvector for cluster $h$ in the
second dimension is $\bis_h = \biX_h^\top\biu_h$. Thus there are $G$
score vectors $\bis_g$ and $H$ score vectors~$\bis_h$. The score
subvector for intersection $gh$ is $\bis_{gh}= \biX_{gh}^\top\biu_{gh}$.

The variance matrix of the scores can always be written as
\begin{equation}
\label{Sigma}
\bSigma = \E (\biX^\top\tn\biu\biu^\top\tn\biX )
= \sum_{g,g'=1}^G \sum_{h,h'=1}^H \E \big(\bis_{gh}\bis_{g'h'}^\top\big).
\end{equation}
Under two\tkk-way clustering, it must be the case that
\begin{equation}
\label{def Omega}
\E (\bis_{gh}\bis_{g'h'}^\top) = \bzero 
\quad \text{if} \quad g' \neq g \textrm{ and } \, h' \neq h,
\end{equation}
but the covariances may be arbitrary when either $g=g'$ or $h=h'$. The
variance matrices for the score subvectors $\bis_g$, $\bis_h$, and
$\bis_{gh}$ are respectively denoted 
\begin{equation}
\label{var matrices}
\bSigma_g = \E(\bis_g \bis_g^\top), \;\;
\bSigma_h = \E(\bis_h \bis_h^\top), \;\text{ and }\;
\bSigma_{gh} = \E(\bis_{gh} \bis_{gh}^\top).
\end{equation}
From \eqref{def Omega} and \eqref{var matrices}, it is evident that
\begin{equation}
\label{truesig}
\bSigma = \sum_{g=1}^G \bSigma_g + \sum_{h=1}^H \bSigma_h 
    - \sum_{g=1}^G\sum_{h=1}^H \bSigma_{gh}.
\end{equation}
This follows from the inclusion-exclusion principle. The third term in
\eqref{truesig} is essential to avoid double\tkk-counting, but, as we
shall see, it causes practical difficulties for estimating~$\bSigma$.

As usual, the OLS estimator of $\bbeta$ is $\hat\bbeta=(\biX^\top
\biX)^{-1} \biX^\top \biy$, and the OLS residual vector is~$\hat\biu$.
The subvectors of $\hat\biu$ for cluster~$g$, cluster~$h$, and the
intersection~$gh$ are denoted $\hat\biu_g$, $\hat\biu_h$, and
$\hat\biu_{gh}$, respectively. From standard arguments for sandwich
variance matrices,
\begin{equation}
\Var(\hat\bbeta) = (\biX^\top\biX)^{-1}\bSigma\tk(\biX^\top\biX)^{-1} =
\biV_G + \biV_H - \biV_I,
\label{Vtrue}
\end{equation}
where the component matrices are
\begin{align}
\label{VG}
\biV_G &=
(\biX^\top\biX)^{-1}\biggl(\,\sum_{g=1}^G\bSigma_g\!\biggr)
(\biX^\top\biX)^{-1},\\
\label{VH}
\biV_H &=
(\biX^\top\biX)^{-1}\biggl(\,\sum_{h=1}^H\bSigma_h\!\biggr)
(\biX^\top\biX)^{-1}, \mbox{ and} \\
\label{VI}
\biV_I &=
(\biX^\top\biX)^{-1}\biggl(\,\sum_{g=1}^G\sum_{h=1}^H\bSigma_{gh}\!\biggr)
(\biX^\top\biX)^{-1}\tn.
\end{align}
The empirical analog of \eqref{Vtrue} is the three\tkk-term two\tkk-way CRVE
\begin{equation}
\label{3mat}
\hat\biV_1^{(3)} = \hat\biV_G + \hat\biV_H - \hat\biV_I,
\end{equation}
where the estimators on the right-hand side of \eqref{3mat} correspond
to \eqref{VG}, \eqref{VH}, and~\eqref{VI} and will be defined shortly.
The subscript ``1'' in $\hat\biV_1^{(3)}$ identifies this as a
CV$_{\tn1}$ estimator, by analogy with the HC$_1$ estimator of
\citet{MW_1985}. The three component estimators in \eqref{3mat} are
based on the empirical score subvectors $\hat\bis_g$, $\hat\bis_h$,
and~$\hat\bis_{gh}$, which take the same form as the actual score
subvectors, but with $\hat\biu$ replacing~$\biu$. Thus they are all
CV$_{\tn1}$ estimators:
\begin{align}
\label{VhatG}
\hat\biV_G &= \frac{G(N-1)}{(G-1)(N-k)}
   (\biX^\top\biX)^{-1}\bigg(\sum_{g=1}^G\hat\bis_g\hat\bis_g^\top\!\bigg)
   (\biX^\top\biX)^{-1},\\
\label{VhatH}
\hat\biV_H &= \frac{H(N-1)}{(H-1)(N-k)}
   (\biX^\top\biX)^{-1}\bigg(\sum_{h=1}^H\hat\bis_h\hat\bis_h^\top\!\bigg)
   (\biX^\top\biX)^{-1},\;\;\mbox{and}\\
\label{VhatI}
\hat\biV_I &= \frac{I(N-1)}{(I-1)(N-k)}
   (\biX^\top\biX)^{-1}\bigg(
   \sum_{g=1}^G\sum_{h=1}^H\hat\bis_{gh}\hat\bis_{gh}^\top\!\bigg)
   (\biX^\top\biX)^{-1}.
\end{align}
The leading scalar factors here are analogous to the scalar factor for
the usual one\tkk-way CRVE. Since some of the intersections may
contain no observations, some of the $\hat\bis_{gh}$ may not exist. In
practice, it may therefore be advisable to replace the double
summation in \eqref{VhatI} with a single summation over all non-empty
intersections.

The superscript ``(3)'' on $\hat\biV_1^{(3)}$ in \eqref{3mat}
emphasizes that this estimator has three terms, which correspond to
the three terms in \eqref{truesig}. Because $\hat\biV_I$ is subtracted
from the sum of $\hat\biV_G$ and $\hat\biV_H$, the matrix
$\hat\biV_1^{(3)}$ is not necessarily positive definite in finite
samples. This problem is not trivial, and there is more than one way
to deal with it.

One approach, suggested in \citet{CGM_2011} and implemented in
\texttt{Stata}, Version 18 and later, is to compute the eigenvalues of
$\hat\biV_1^{(3)}$, say $\lambda_1,\ldots,\lambda_k$. When any of them
is not positive, $\hat\biV_1^{(3)}$ is replaced by the
eigen-decomposition $\hat\biV_1^{(3+)} =
\biU\bLambda^{\!+}\biU^\top$\tn\tn, where $\biU$ is the $k\times k$
matrix of eigenvectors and $\bLambda^{\!+}$ is a diagonal matrix with
typical diagonal element $\lambda_j^+ = \max \{ \lambda_j,0 \}$. In
practice, it may be numerically safer to compare the eigenvalues with
a very small positive number, say $\eta$, and define $\lambda_j^+$ as
$\max \{\lambda_j,\eta\}$. In our programs, we use $\eta=10^{-12}$.
Doing this ensures that $\hat\biV_1^{(3+)}$ is positive definite,
albeit just barely so.

This approach is not entirely satisfactory. Wald statistics and
$t$-statistics based on $\hat\biV_1^{(3+)}$ are computable, but they
may be extremely large. Even when this does not happen, and all
quantities of interest can be computed using~$\hat\biV_1^{(3)}$,
replacing $\hat\biV_1^{(3)}$ by $\hat\biV_1^{(3+)}$ can change all the
standard errors. Moreover, the standard error of any element
of~$\hat\bbeta$, say~$\hat\beta_j$, is not invariant to nonsingular
transformations of the remaining columns of the matrix~$\biX$\tn.
Thus, for example, precisely how fixed effects or other dummy
variables are specified may affect the standard error
of~$\hat\beta_j$, even though $\hat\beta_j$ itself is invariant to
such reparametrizations. For instance, if one wanted to control for 
American state fixed effects, the choice of using either Texas or 
California as the reference group can change the estimated standard 
error for the treatment regressor of interest.

A simpler way to avoid the problem that $\hat\biV_1^{(3)}$ may not be
positive definite is to replace it by the two\tkk-term estimator
\begin{equation}
\label{2mat crve}
\hat\biV_1^{(2)} = \hat\biV_G + \hat\biV_H.
\end{equation}
This estimator has been studied in \citet{Davezies_2018}. It omits the
third term in \eqref{3mat} and therefore involves double\tkk-counting.
The justification for omitting $\hat\biV_I$ is that, under additional 
regularity conditions, it becomes asymptotically negligible as both
$G$ and $H$ tend to infinity. Because
\begin{equation}
\label{V3-V2}
\hat\biV_1^{(2)} - \hat\biV_1^{(3)} = \hat\biV_I
\end{equation}
is positive definite, it follows that a Wald statistic or
$t$-statistic based on $\hat\biV_1^{(2)}$ will always be smaller than
the same statistic based on~$\hat\biV_1^{(3)}$, so that the former is 
more conservative.

The conditions for consistency of $\hat\biV_1^{(2)}$ are stronger
than the ones needed for~$\hat\biV_1^{(3)}$. For example,
\citet{MNW_2021} shows that whenever the scores are actually
independent, or whenever they are only correlated at the intersection
level, $\hat\biV_1^{(2)}$ yields test statistics that are
asymptotically too small. In this case, $\hat\biV_G \approx 
\hat\biV_H \approx \hat\biV_I$. Therefore,
\begin{equation}
\label{V2hatx}
\hat\biV_1^{(2)} = \hat\biV_G + \hat\biV_H \approx 2\hat\biV_I,
\end{equation}
whereas
\begin{equation}
\label{V3hatx}
\hat\biV_1^{(3)} = \hat\biV_G + \hat\biV_H - \hat\biV_I
\approx \hat\biV_I.
\end{equation}
Thus, in this case, $\hat\biV_1^{(2)}$ is approximately twice as large
as $\hat\biV_1^{(3)}$, and twice as large as it should be. The use of
``$\approx$'' in \eqref{V2hatx} and \eqref{V3hatx} is deliberately
informal, since we did not take limits or introduce any factors of the
sample size in~\eqref{Vtrue}. For a rigorous treatment, see
\citet[Theorem~1]{MNW_2021}. The result \eqref{V2hatx} suggests that
$\hat\biV_1^{(2)}$ is also likely to perform poorly in finite samples
when most of the intra-cluster correlation is at the intersection
level. 

We now propose a third way to avoid cases in which test statistics
based on the three\tkk-term estimator $\hat\biV_1^{(3)}$ are not
positive. Our proposal is simply to compute three test statistics and
use the one that takes the smallest positive value. For the hypothesis
that $\biR\bbeta=\bir$, the three Wald statistics are
\begin{equation}
\begin{aligned}
\label{Wald3}
W_3 &= (\biR\tkk\hat\bbeta - \bir)^\top (\biR\tk\hat\biV_1^{(3)}
\biR^\top)^{-1}(\biR\tkk\hat\bbeta - \bir),\\
W_G &= (\biR\tkk\hat\bbeta - \bir)^\top (\biR\hat\biV_G\biR^\top)^{-1}
(\biR\tkk\hat\bbeta - \bir),\;\mbox{and}\\
W_H &= (\biR\tkk\hat\bbeta - \bir)^\top (\biR\hat\biV_H\biR^\top)^{-1}
(\biR\tkk\hat\bbeta - \bir).
\end{aligned}
\end{equation}
The statistic we propose to use is
\begin{equation}
\label{Wmin}
W_{\min} = \min \big\{{\rm pos}(W_3), W_G, W_H \big\} ,
\end{equation}
where ${\rm pos}(W_3)$ equals~$W_3$ whenever $W_3$ is positive
and~$+\infty$ whenever it is either negative or undefined, as it can
be when $\hat\biV_1^{(3)}$ is not positive definite. By using
$W_{\min}$ defined in \eqref{Wmin}, we not only avoid Wald statistics
that are not positive numbers but also Wald statistics that are
misleadingly large. In a particular sample, one or more diagonal
elements of $\hat\biV_I$ may randomly happen to be just a little
smaller than the sum of the corresponding elements of $\hat\biV_G$
and~$\hat\biV_H$. Thus $\hat\biV_1^{(3)}$ can yield extremely large
test statistics which are completely misleading.

In most cases, it is not necessary to calculate the entire
$\hat\biV_1^{(3)}$ matrix. Only the rows and columns needed for the
Wald statistic have to be calculated. Of course, when there is only
one restriction, we can use a $t$-statistic instead of a Wald
statistic. In this case, we just need to find the largest of the three
standard errors and calculate a $t$-statistic using that standard
error. We will refer to our procedure as the ``max-se'' procedure
because the case of just one restriction is by far the most common
one. The max-se procedure has recently been studied in
\citet{Davezies_2025}, which cites an earlier version of this paper.

Henceforth, we denote the variance and standard error estimators based
on $\hat\biV_1^{(2)}$ and $\hat\biV_1^{(3)}$ as CV$_{\tn1}^{\tk(2)}$ and
CV$_{\tn1}^{\tk(3)}$ estimators, respectively, the ones based on
$\hat\biV_1^{(3+)}$ as CV$_{\tn1}^{\tk(3+)}$ estimators, and the ones
implicit in \eqref{Wmin} as CV$_{\tn1}^{\tk(\max)}$ estimators. In the
scalar case, $\text{CV}_{\tn1}^{\tk(\max)}$ is $\hat V_1^{(\max)} = \max
\{\hat V_1^{(3)}, \hat V_G , \hat V_H \}$. This explains the 
``$(\max )$'' superscript and also makes it clear that, asymptotically, 
the CV$_{\tn1}^{\tk(3)}$, CV$_{\tn1}^{\tk(3+)}$, and 
CV$_{\tn1}^{\tk(\max)}$ estimators must be identical whenever the 
scores are positively correlated in either or both of the $G$ and $H$ 
dimensions. 

In most cases where it makes sense to specify $\bSigma$ as 
in~\eqref{Sigma}, the CV$_{\tn1}^{\tk(3)}$, CV$_{\tn1}^{\tk(3+)}$, and
CV$_{\tn1}^{\tk(\max)}$ estimators will have exactly the same asymptotic
properties. They may or may not be identical in practice. In fact,
there are cases where they may differ greatly. This seems to be most
common when there is very little intra-cluster correlation and/or the
number of clusters is small, and/or the number of regressors is large,
as we shall see in \Cref{sec:simuls}.

\section{Two\tkk-\tn Way Cluster-Jackknife CRVEs}
\label{sec:twojack}

The component CRVEs defined in \eqref{VhatG}, \eqref{VhatH}, and
\eqref{VhatI} all have the form of the widely-used CV$_{\tn1}$
estimator. However, recent work by \citet{MNW-bootknife} and
\citet{Hansen-jack} strongly suggests that, in the one\tkk-way case,
it is better to use a CRVE based on the cluster jackknife, which is
analogous to the HC$_3$ estimator of \citet{MW_1985}. The key idea of
the cluster jackknife is to compute $G$ (or~$H$ or~$I$) sets of
parameter estimates, each of which omits one cluster at a time, and
then compute a CRVE using the variation among these estimates.

Let $J \in \{ G, H, I \}$, and let $j$ denote the corresponding
lower-case letter. In the intersection dimension ($J=I$), the summation
$\sum_{j=1}^J \biZ_j$ should be interpreted as $\sum_{g=1}^G\sum_{h=1}^H
\biZ_{gh}$ for any $\biZ$, where $I$ denotes the number of non-empty 
intersections, which may be smaller than~$GH$. The OLS estimates of 
$\bbeta$ when each cluster in the $J$ dimension is omitted in turn are
\begin{equation}
\label{delone}
\hat\bbeta^{(j)} = (\biX^\top\!\biX - \biX_j^\top\!\biX_j)^{-1}
(\biX^\top\biy - \biX_j^\top\biy_j), \quad j=1,\ldots,J.
\end{equation}
Then the component cluster-jackknife variance matrix estimators are
\begin{equation}
\hat\biV_J^{\rm JK} =
\frac{J-1}{J} \sum_{j=1}^J (\hat\bbeta^{(j)} -
\hat\bbeta)(\hat\bbeta^{(j)} - \hat\bbeta)^\top
\quad \mbox{for } \{ j,J \} = \{ g,G \} , \{ h,H \} , \{ i,I \} .
\label{jackj}
\end{equation}
Thus the three\tkk-term jackknife CRVE is
\begin{equation}
\label{3jack}
\hat\biV_3^{(3)} = \hat\biV_G^{\rm JK} + \hat\biV_H^{\rm JK} -
\hat\biV_I^{\rm JK},
\end{equation}
which is analogous to \eqref{3mat}. The subscript ``3'' here follows
the usual notation for jackknife variance matrices; see
\citet{MNW-bootknife}. There is also a two\tkk-term jackknife CRVE
and, more interestingly, one that is analogous to the
CV$_{\tn1}^{\tk(\max)}$ estimator. We refer to the three CRVEs based
on the cluster jackknife as CV$_{\tn3}^{\tk(2)}$,
CV$_{\tn3}^{\tk(3)}$, and~CV$_{\tn3}^{\tk(\max)}$.

The CRVEs defined in \eqref{jackj} are not the only cluster-jackknife
variance matrix estimators. Instead of computing variances around
$\hat\bbeta$, one can instead compute them around (in the two\tkk-way
case) the three sample averages, $\bar\bbeta^J = J^{-1}\sum_{j=1}^J
\hat\bbeta^{(j)}$. This makes the alternative CRVEs a little smaller
than the ones given in \eqref{jackj}. Because simulation experiments
in \citet{BM_2002} and \citet{MNW-bootknife} suggest that, in the
one\tkk-way case with $G$ clusters, inferences based on the
alternative jackknife CRVE are almost identical to ones based on
$\hat\biV_G^{\rm JK}$\tn, we do not study the former in this paper.

Computing the component CRVEs in \eqref{jackj} that are needed for
CV$_{\tn3}^{\tk(3)}$, CV$_{\tn3}^{\tk(3+)}$\tn, and
CV$_{\tn3}^{\tk(\max)}$ is somewhat more work than computing the ones
in \eqref{VhatG}, \eqref{VhatH}, and \eqref{VhatI} that are needed for
CV$_{\tn1}^{\tk(3)}$, CV$_{\tn1}^{\tk(3+)}$\tn, and
CV$_{\tn1}^{\tk(\max)}$, especially when the number of non-empty
intersections, $I$\tn, is large. The first thing is to calculate the
cluster-level matrices and vectors
\begin{equation}
\biX_j^\top\!\biX_j \;\;\mbox{and}\;\; \biX_j^\top\biy_j, \quad
j=1,\ldots,J,
\quad \mbox{for } \{ j,J \} = \{ g,G \} , \{ h,H \} , \{ i,I \} .
\label{subthings}
\end{equation}
These quantities can be computed for the intersections with a single
pass over the $N$ observations. The ones for the $G$ and $H$
dimensions are just summations of the ones for the appropriate
intersections. The three sets of $\hat\bbeta^{(j)}$ can then be
computed using \eqref{delone} for the three clustering dimensions.
Unfortunately, this may be expensive when both $k$ and $I$ are large,
because computing the omit-\tkk one\tkk-cluster estimates for the
intersections involves inverting $I$ different $k\times k$ matrices.

When computational cost is a concern, it can be reduced significantly 
by replacing $\hat\biV_I^{\rm JK}$ in~\eqref{3jack} with~$\hat\biV_I$, 
yielding the mixed three-term estimator
\begin{equation}
\label{31jack}
\hat\biV_{3,1}^{(3)} = \hat\biV_G^{\rm JK} + \hat\biV_H^{\rm JK} -
\hat\biV_I.
\end{equation}
Because $\hat\biV_I$ is almost always smaller than $\hat\biV_I^{\rm
JK}$, $\hat\biV_{3,1}^{(3)}$ will generally be larger than
$\hat\biV_3^{(3)}$. However, unless $I$ is small (which can only
happen if both $G$ and $H$ are small or $I$ is much smaller 
than~$GH$), the matrices $\hat\biV_I$ and $\hat\biV_I^{\rm JK}$ tend to 
be very similar. Thus the difference between \eqref{3jack} and
\eqref{31jack} is negligible in most cases, as discussed near the end
of \Cref{subsec:clustsize}. However, it can be noticeable when there
are many empty intersections; see \Cref{subsec:empty}.

In many cases, the regression model \eqref{model} will include fixed
effects in the $G$ and $H$ dimensions; that is, two\tkk-way fixed
effects. If so, it may be rewritten as
\begin{equation}
\label{TWFE}
\biy = \biZ\bbeta_p + \biD^G\bgamma + \biD^H\bdelta + \biu.
\end{equation}
Here the matrix $\biZ$, which has $p$ columns, corresponds to the
actual explanatory variables, and $\bbeta_p$ contains the elements of
$\bbeta$ for those variables. The matrices $\biD^G$ and $\biD^H$
contain dummy variables for the fixed effects in dimensions $G$
and~$H$, respectively. Collectively, these have $G+H-1$ columns, say
$G$ for $\biD^G$ and $H-1$ for~$\biD^H$\tn. Thus
$\biX=[\biZ\;\;\biD^G\;\;\biD^H]$, and $k=p+G+H-1$.

For the model \eqref{TWFE}, there is an important computational issue.
It is impossible to invert the matrices $\biX^\top\biX - 
\biX_g^\top\biX_g$ and $\biX^\top\biX - \biX_h^\top\biX_h$ in
\eqref{subthings}, because for each of them the row and column
corresponding to the fixed effect for cluster $g$ or cluster $h$
contains only zeros. There are three ways to deal with this issue. The
first is just to drop the subsamples in which the inversion is not 
possible. This is the default in many standard software routines, such
as the prefix \texttt{jackknife} in \texttt{Stata}. However, it is not
viable for~\eqref{TWFE}, because every jackknife replication would
have to be dropped. The second approach is to replace the inverse in
\eqref{delone} by a generalized inverse. Then all of the coefficients
except the fixed effect for the omitted cluster can be computed, and
the latter is set to zero. Thus, whenever there are two\tkk-way fixed
effects, $\hat\biV_3^{(3)}$ in \eqref{3jack} is effectively being
defined as a $p\times p$ variance matrix for~$\hat\bbeta_p$ instead of
a $k\times k$ variance matrix for~$\hat\bbeta$.

The third approach is to to partial out the cluster fixed effects
before computing the one\tkk-way CRVEs. However, this must be done
with great care. It is valid to partial out cluster fixed effects in
the $G$ dimension when calculating $\hat\biV_G^{\rm JK}$\tn, but it is
invalid to partial them out when calculating either $\hat\biV_H^{\rm
JK}$ or~$\hat\biV_I^{\rm JK}$\tn. The problem is that, after the
cluster fixed effects in the $G$ dimension have been partialed out,
the observations for every cluster in the $H$ and $I$ dimensions
generally depend on observations in some or all of the other clusters
in those dimensions. Thus $\hat\bbeta^{(h)}$ and $\hat\bbeta^{(i)}$
would not actually be vectors of omit-\tkk one\tkk-cluster estimates.
Similarly, it is invalid to partial out fixed effects in the $H$
dimension when calculating either $\hat\biV_G^{\rm JK}$ or
$\hat\biV_I^{\rm JK}$\tn. The $I$ dimension is always the most
expensive one to deal with, because it involves the largest number of
clusters, and it is not valid to partial out fixed effects in either
the $G$ or $H$ dimensions when calculating~$\hat\biV_I^{\rm JK}$\tn.
This makes it particularly attractive to use \eqref{31jack} instead of
\eqref{3jack} when there are two\tkk-way fixed effects.

It is conventional to employ the Student's $t$ distribution with $\min
\{ G,H \} -1$ degrees of freedom to obtain $P$ values or critical
values for $t$-statistics based on~CV$_{\tn1}^{\tk(3)}$. As in the
one\tkk-way case, it seems reasonable to use the same distribution for
$t$-statistics based on CV$_{\tn3}^{\tk(3)}$ as well, and this is the
approach that we take.

However, at least two other methods could in principle be used. For
one\tk-way clustering, \citet{BM_2002} proposes a way to obtain
approximate critical values for \mbox{$t$-tests} based on CV$_{\tn1}$
by using a $t$ distribution with a calculated degrees\tkk-of-freedom
parameter; see also \citet{Imbens_2016}. Another method based on the
same idea is proposed in \citet{Hansen-jack,Hansen_2025}. For the
two\tkk-way case, one could in principle use the same sort of
approximate critical value. However, we are not aware of any method
for obtaining such a critical value for $t$-statistics based on
two\tkk-way clustering. This is an area for future research.

Another possibility is to use bootstrap methods. The pigeonhole
bootstrap of \citet{Owen_2007} was studied in \citet{Menzel_2021} and
found to be conservative in general. That paper also proposed some new
and rather complicated bootstrap procedures for inference about the
sample mean. The wild cluster bootstrap \citep*{CGM_2008,DMN_2019} has
been widely used for inference with one\tkk-way clustering, and
\citet{MNW_2021} suggested using it for two\tkk-way clustering as
well. In that paper, the usual wild cluster bootstrap for one of the
$G$, $H$, or $I$ dimensions is used to generate the bootstrap samples.
This procedure is not entirely satisfactory, because the bootstrap
samples cannot reproduce the intra-cluster covariances among the
residuals. Nevertheless, this wild bootstrap routine is conveniently
and efficiently coded in the \texttt{boottest} package in
\texttt{Stata} using the \texttt{bootclust} option; see \citet*{RMNW}
for details. Recently, \citet{Hounyo-boot} proposes a wild bootstrap
DGP that gives positive weight to both dimensions.

In the absence of any satisfactory alternative, we currently recommend
using the cluster jackknife together with critical values based on the
Student's $t$ distribution with $\min\{ G,H \} -1$ degrees of freedom.
As we shall see in \Cref{sec:simuls}, this approach often works 
remarkably well. Whether combining the jackknife with a bootstrap 
procedure would perform even better is a topic for future research;
see \citet{MNW-bootknife} for evidence on this with one\tkk-way 
clustering.

Computing the three\tkk-term cluster-jackknife estimator for the
two\tkk-way fixed-effects model \eqref{TWFE} can be costly when $G$
and $H$ are not fairly small. The cost of forming the
$\biX_j^\top\biX_j$ matrices and the $\biX_j^\top\biy_j$ vectors is
roughly $O(Nk^2) = O(N(G+H+p-1)^2)$, because $\biX$ has $k = p+G+H-1$
columns. Since \eqref{delone} has to be computed $G+H+I \approx
G+H+GH$ times, the cost of computing the cluster-jackknife estimates
after the $\biX_j^\top\biX_j$ matrices and $\biX_j^\top\biy_j$ vectors
have been formed is roughly $O(GHk^2)=O(GH(G+H+p-1)^2)=O(G^4)$ if
$G\approx H$\tn.

Most of the computational cost of the two\tkk-way cluster jackknife
arises from the need to deal with the $I \le GH$ intersections. When
$I<\!<GH$, the cost can be greatly reduced if the empty intersections
are skipped when calculating the omit-\tkk one\tkk-cluster estimates
using~\eqref{delone}. An additional reduction is possible by using
\eqref{31jack} instead of \eqref{3jack}.

\section{Properties of the Cluster-Jackknife CRVE}
\label{sec:properties}

Properties of classic-jackknife variance estimators are well known. 
However, for the cluster jackknife, the only analysis of theoretical 
properties that we are aware of is in \citet{Hansen-jack}. In the
context of the linear regression model with one\tkk-way clustering, it
shows that a certain cluster-jackknife variance estimator (which is
not quite the same as $\hat\biV_3$, but should usually be very
similar) is never downward biased. Moreover, the associated $t$-tests
and confidence intervals have worst-case size, or coverage, that is
controlled by the Cauchy distribution. In contrast, variance
estimators based on CV$_{\tn1}$ can be severely downward biased, the
associated $t$-tests have worst-case size of~1, and the associated
confidence intervals have worst-case coverage of~0.

\subsection{Consistency of the Cluster-Jackknife CRVE}
\label{subsec:consistency}

In this subsection, we prove consistency of the two\tkk-way cluster 
jackknife~CRVE. We will need the following two assumptions.

\begin{assumption}
\label{assn:rank}
Let $J \in \{ G, H, I \}$, and let $j$ denote the corresponding
lower-case letter. The omit-\tkk one\tkk-cluster matrices, 
$\biX^\top\biX - \biX_j^\top\biX_j$, are invertible for all 
$j=1,\ldots,J$.
\end{assumption}

\begin{assumption}
\label{assn:yap}
Let subscript $i=1,\ldots,N$ denote an observation, and let the smallest 
eigenvalue of~$\bSigma$ in~\eqref{Sigma} be denoted $\lambda_N 
= \lambda_{\min}(\bSigma)$. There exists $K_0<\infty$ and $K_1>0$ such 
that:
\begin{itemize}
\item[(a)] $\E ( u_i^4 |\biX_i )\leq K_0$, $\E ( \Vert \biX_i \Vert ^4 ) 
\leq K_0$, and $\E (\bis_i =0)$ for all $i=1,\ldots,N$.
\item[(b)] $\lambda_N^{-1} \max_{g=1,\ldots,G}N_g^2 \to 0$ and 
$\lambda_N^{-1} \max_{h=1,\ldots,H}N_h^2 \to 0$.
\item[(c)] $\lambda_N^{-1} \sum_{g=1}^G N_g^2 \leq K_0$, and 
$\lambda_N^{-1} \sum_{h=1}^H N_h^2 \leq K_0$.
\item[(d)] If observations~$i$ and~$j$ do not share a cluster in either 
dimension, then $(\biX^\top_i\!, u_i)$ is independent 
of~$(\biX^\top_j\!, u_j)$.
If observations~$i_1,i_2$ share a cluster, observations~$j_1,j_2$ 
share a cluster, but neither $i_1$ nor~$i_2$ share a cluster with $j_1$ 
or~$j_2$, then $(\biX^\top_{i_1}, u_{i_1} , \biX^\top_{i_2}, u_{i_2})$ 
is independent of~$(\biX^\top_{j_1},u_{j_1},\biX^\top_{j_2},u_{j_2})$.
\item[(e)] $\lambda_{\min}( N^{-1} \E ( \biX^\top \biX ))\geq K_1$.
\end{itemize}
\end{assumption}

\Cref{assn:rank} guarantees existence of the omit-\tkk one\tkk-cluster
estimators in~\eqref{delone}, and hence the cluster jackknife. Note
that this rules out cluster fixed effects and other cases where a 
regressor is non-zero for only one cluster (in any dimension).
Practical ways to deal with this situation were discussed in the
previous section. Existing proofs for the validity of other
two\tkk-way CRVEs, such as \citet{Davezies_2025} and \citet{yap_2025},
also rule out cluster fixed effects.

\Cref{assn:yap} is identical to Assumption~3 in \citet{yap_2025}, but 
adapted to our notation. Part~(a) is a standard moment condition. 
Parts~(b) and~(c) restrict the heterogeneity of cluster sizes and rule 
out degenerate non-Gaussian cases analyzed in \citet{Menzel_2021}. 
They can be viewed as generalizations of Assumptions~2 and~3 in 
\citet{DMN_2019} to allow two\tkk-way clustering. Part~(d) strengthens 
\eqref{def Omega} to independence, and part~(e) is a version of the 
usual rank condition for~OLS. See \citet{yap_2025} for a detailed 
discussion. An alternative asymptotic framework is that of separately 
exchangeable arrays \citep[e.g.,][and others]{Davezies_2018}, under 
which our results could also be proven by the same arguments 
\citep[Proposition~1]{yap_2025}.

\begin{theorem}
\label{thm:cons} 
Under \Cref{assn:rank,assn:yap} the following hold.
\begin{itemize}
\item[(i)] $(\Var(\hat\bbeta))^{-1}\hat\biV_3^{(3)} \Pto 1$,
$(\Var(\hat\bbeta))^{-1}\hat\biV_3^{(3+)} \Pto 1$, and 
$(\Var(\hat\bbeta))^{-1}\hat\biV_{3,1}^{(3)} \Pto 1$. 
\item[(ii)] For a conforming vector $\bia \neq \bzero$, if
$(\Var(\bia^\top\hat\bbeta ))^{-1}\max \{ \bia^\top\hat\biV_1^{(3)}\bia
, \bia^\top\hat\biV_G\bia , \bia^\top\hat\biV_H\bia \} \Pto 1$, then
$(\Var(\bia^\top\hat\bbeta ))^{-1}\max \{ \bia^\top\hat\biV_3^{(3)}\bia 
, \bia^\top\hat\biV_G^{\rm JK}\bia , \bia^\top\hat\biV_H^{\rm JK}\bia \}
\Pto 1$.
\item[(iii)] If $\biV_I$ is asymptotically negligible in 
\eqref{Vtrue}, then $(\Var(\hat\bbeta))^{-1}\hat\biV_3^{(2)} \Pto 1$.
\end{itemize}
\end{theorem}

A proof of \Cref{thm:cons} is given in \Cref{app:proof}. Under 
\Cref{assn:yap}, \citet[Proposition~2]{yap_2025} shows that 
$( \Var ( \hat\bbeta ))^{-1/2}(\hat\bbeta - \bbeta_0 ) \dto \N(\bzero , 
\bfI_k)$. 
Combined with our \Cref{thm:cons}(i), this implies that $t$-tests or 
$F$-tests based on the three jackknife CRVEs, CV$_3^{(3)}$, 
CV$_3^{(3+)}$, and CV$_{3,1}^{(3)}$, have correct asymptotic size, 
and associated confidence intervals have correct asymptotic coverage.

The results in \Cref{thm:cons}(ii),(iii) require additional conditions 
to obtain asymptotically valid inference. For part~(ii), we have stated
the result for linear contrasts. Sufficient primitive conditions for 
this part were studied in \citet{Davezies_2025}, and under such 
conditions it also holds by \Cref{thm:cons}(ii) that inference based on 
the jackknife CV$_3^{(\max)}$ is asymptotically valid, at least for 
linear contrasts. For part~(iii), $\biV_I$ is asymptotically negligible 
when there is sufficient intra-cluster correlation in either of the two 
main clustering dimensions, $G$ or~$H$ \citep{Davezies_2018,MNW_2021}, 
and then \Cref{thm:cons}(iii) shows that inference based on the 
jackknife CV$_3^{(2)}$ is asymptotically valid. On the other hand, for 
example, if there is clustering only at the intersection level, then 
$\biV_I$ is not asymptotically negligible, and $\hat\biV_1^{(2)}$ is not
consistent; see our discussion around \eqref{2mat crve}--\eqref{V3hatx}.
In that case, $\hat\biV_3^{(2)}$ is not consistent either.

\subsection{Robustness of the Cluster-Jackknife CRVE}
\label{subsec:whyjack}

Simulation results \citep[e.g.,][]{MNW-guide,MNW-bootknife,%
MNW-influence,Hansen-jack,Hansen_2025} have shown that one\tkk-way
cluster-jackknife CRVEs perform better in finite samples than 
conventional CRVEs in terms of coverage of confidence intervals and 
size of tests. In this subsection, we discuss possible reasons 
underlying those results in the context of two\tkk-way clustering. The
key reason seems to be that cluster-jackknife CRVEs handle cluster
size variation, and heterogeneity more generally, better than do
conventional CRVEs. This is particularly important for three\tkk-term
estimators, as we explain.

It is known that CV$_{\tn3}$ estimators are less (downward) biased
than CV$_{\tn1}$ ones \citep[e.g.,][]{Efron-Stein,Hansen-jack}. The
reason for this can be seen intuitively as follows. The one\tkk-way
cluster-jackknife CRVEs in \eqref{jackj} can be rewritten as
\begin{equation}
\label{jackjMgg}
\hat\biV_J^{\rm JK} = \frac{J-1}{J} (\biX^\top\biX )^{-1} \bigg( 
\sum_{j=1}^J \ddot\bis_j\ddot\bis_j^\top \bigg)  (\biX^\top\biX )^{-1}
\quad \mbox{for } \{ j,J \} = \{ g,G \} , \{ h,H \} , \{ i,I \} ,
\end{equation} 
where the modified score vectors $\ddot\bis_j$ are defined as
\begin{equation}
\label{ddotbis}
\ddot\bis_j = \biX^\top\!\biM_{jj}^{-1} \hat\biu_j,
\end{equation}
and $\biM_{jj}$ denotes the \th{(j,j)} block of~$\biM_{\biX} = \bfI_N
- \biX (\biX^\top\biX )^{-1}\biX^\top$. For a proof of equality of 
\eqref{jackj} and \eqref{jackjMgg}, see 
\citet[pp.~675--676]{MNW-bootknife}. Normalizing the modified score
vectors in \eqref{ddotbis} by the factor $\biM_{jj}^{-1}$ undoes some
of the shrinkage caused by least squares. Since the $\biM_{jj}$ are
inversely related to cluster leverage \citep{MNW-influence}, the
cluster-jackknife CRVE puts more weight on clusters with high leverage
compared with the CV$_1$ estimator. This accounts for the smaller bias
of the former relative to the latter, because high-leverage clusters
are relatively more important in determining the actual variance of
the estimator.

In many two\tkk-way designs, clusters vary greatly in size and/or
leverage in one or both dimensions, or there are few clusters in one 
dimension. Thus, one or both of $\hat\biV_G$ and $\hat\biV_H$ is likely
to be seriously downward biased. However, $\hat\biV_I$ is usually based
on a much larger number of clusters, often as many as $G \times H$. As 
a result, its downward bias is likely to be comparatively moderate. In
consequence, when $\hat\biV_I$ is subtracted from the sum of
$\hat\biV_G$ and $\hat\biV_H$ to form $\hat\biV_{\tn1}^{(3)}$\tn,
there is a good chance that the latter will be very severely biased.
In contrast, the arguments above and results in \citet{Hansen-jack} 
(for one\tkk-way clustering) suggest that $\hat\biV_G^{\rm JK}$ and 
$\hat\biV_H^{\rm JK}$ are never downward biased, although either or both
may be upward biased. It is possible that $\hat\biV_I^{\rm JK}$ may be upward
biased in this case, but since it is normally based on a much larger
number of clusters, any such bias is likely to be modest, and
subtracting it is not likely to cause much downward bias in
$\hat\biV_{\tn3}^{(3)}$ itself. These arguments suggest that
$\hat\biV_{\tn3}^{(3)}$ is more likely to be positive definite than
$\hat\biV_{\tn1}^{(3)}$ and that tests based on
$\hat\biV_{\tn3}^{(3)}$ should be more reliable than ones based
on~$\hat\biV_{\tn1}^{(3)}$.

The above arguments imply that, if the sample is heterogeneous in only
one dimension, so that only one of $\hat\biV_G$ and $\hat\biV_H$ is
severely downward biased, then the downward bias in
$\hat\biV_{\tn1}^{(3)}$ is likely to be relatively moderate. This case
probably occurs quite often in panel settings, where samples (and 
cluster sizes) are often heterogeneous across cross-sectional units
but homogeneous across time periods. For instance, a commonly used 
dataset like the Current Population Survey (CPS) will be unbalanced in 
terms of the number of observations per state, but strongly balanced in 
terms of the number of observations per year. In such cases, we would 
still expect CV$_{\tn3}$-based estimators to be more accurate than
CV$_{\tn1}$-based ones, but probably by a smaller margin than in cases
with double heterogeneity.

In empirical research, it is very commonly found that some
intersections of the two clustering dimensions contain no
observations. The possibility of empty intersections can be important
for two\tkk-way clustering, but it cannot arise for one\tkk-way
clustering. To examine the importance of empty intersections, consider
two hypothetical samples, each with $G=H=10$. Thus there are $100$
intersections. For one sample, no intersections are empty, but 70 of
them contain just 1 observation. For the other sample, there are 70
empty intersections. Now consider the cluster-jackknife estimator,
$\hat\biV_I^{\rm JK}$. In the first sample, it is based on 100 terms.
Since dropping just one observation should not change
$\hat\bbeta^{(i)}$ very much, the terms in the summation in 
\eqref{jackj} corresponding to the tiny intersections must all be very
small. In the second sample, the cluster-jackknife estimate is based
on just 30 terms. The terms that were small in the first sample have 
vanished, which seems to be a small difference. The only other 
difference between the two samples is that the leading factor in 
$\hat\biV_I^{\rm JK}$ will be 99/100 in the first sample and 29/30 in
the second, which seems inconsequential. Thus the cluster-jackknife 
estimator handles empty intersections in a reasonable fashion.

\section{Simulation Experiments}
\label{sec:simuls}

Most of our experiments deal with the two\tkk-way fixed-effects
model~\eqref{TWFE}. The number of coefficients is $k=p+G+H-1$, but we
focus on tests of a single coefficient, say~$\beta_1$. Although
\eqref{TWFE} is very widely used, many existing simulation experiments
for two\tkk-way clustering do not include cluster fixed effects. This
is probably because, when the intra-cluster correlations are generated
by a random-effects model, cluster fixed effects absorb all of them.
For example, the experiments in \citet[Section~3.1]{CGM_2011} and
\citet{MNW_2021} do not include fixed effects. In contrast, the
placebo\tkk-regression experiments in Section~3.2 of the former paper
use actual data instead of a random-effects model, and they do include
two\tkk-way fixed effects.

In order to generate data for the model \eqref{TWFE}, the disturbances
must be generated in a way that allows for two\tkk-way intra-cluster
correlation that is not removed by cluster fixed effects. We use
factor models of the form
\begin{equation}
\label{facDGP}
\begin{aligned}
z_{ghi} &= \sigma_g\tk\xi^1_g + \sigma_h\tk\xi^1_h + \sigma_\epsilon
\tk\zeta_{ghi} \;\;\mbox{ if $i$ is odd,}\\
z_{ghi} &= \sigma_g\tk\xi^2_g + \sigma_h\tk\xi^2_h + \sigma_\epsilon
\tk\zeta_{ghi} \;\;\mbox{ if $i$ is even.}
\end{aligned}
\end{equation}
Here $\xi^1_g$ and $\xi^2_g$ are random effects, distributed as
$\N(0,1)$, which apply respectively to the odd-numbered and
even-numbered observations within the \th{g} cluster in the $G$
dimension. Similarly, $\xi^1_h$ and $\xi^2_h$ are $\N(0,1)$ random
effects which apply to the odd-numbered and even-numbered observations
within the \th{h} cluster in the $H$ dimension. The $\zeta_{ghi}$ are
independent standard normals.

The values of $\sigma_g$, $\sigma_h$, and $\sigma_\epsilon$ determine
the amount of correlation for the odd-numbered and even-numbered
observations within each cluster, and hence the correlations within
and across the clusters in the $G$, $H$, and $I$ dimensions. There
will be no correlation for observations that belong to different
clusters in the $G$ and $H$ dimensions. Specifically, the intra-cluster
correlations are $\rho_g$ and $\rho_h$, with $\rho_j=\sigma_j^2$ for
$j=g,h$. To ensure that the $z_{ghi}$ have variance unity, the value
of $\sigma_\epsilon^2 = 1 - \sigma_g^2 - \sigma_h^2$. This
constrains $\rho_g$ and $\rho_h$ not to be too large.

The factor model \eqref{facDGP} provides a simple way to generate data
for a model with two\tkk-way fixed effects. It is based on a
one\tkk-way DGP used in \citet*{MNW-testing} and can be interpreted in
a variety of ways, depending on the nature of the data. The idea is
that there are two types of observations within each cluster in each
dimension, and all the intra-cluster correlation is within each type.
For example, with clustering at the geographical level, there might be
two sub\tkk-regions. With clustering at the industry level, there
might be two types of firm. The key assumption is that the researcher
knows which cluster an observation belongs to in each dimension, but
not which type. Including cluster fixed effects explains some of the
intra-cluster correlation by estimating averages of $\xi^1_g$ and
$\xi^2_g$ for each $G$ cluster and averages of $\xi^1_h$ and $\xi^2_h$
for each $H$ cluster, but it does not explain all of it. Thus
cluster-robust inference is still needed.

In several of the experiments, we focus on cluster size variation.
Following \citet{MW-JAE} and \citet{DMN_2019}, the cluster sizes in the
$G$ dimension are given by
\begin{equation}
N_g = \left[N \frac{\exp(\gamma g/G)}{\sum_{j=1}^G \exp(\gamma
j/G)}\right]\!, \quad g=1,\ldots, G-1,
\label{gameq}
\end{equation}
where $[x]$ denotes the integer part of~$x$. The value of $N_G$ is then
set to $N - \sum_{g=1}^{G-1} N_g$. The formula \eqref{gameq}, perhaps
with a different value of $\gamma$, is also used in the $H$ dimension.
Assuming that the distributions are independent, $N_{gh} \approx N_g
N_h / N$.  In a final step, the cluster sizes are adjusted to ensure
that they are all integers with $N = \sum_{g=1}^G N_g = \sum_{h=1}^H
N_h = \sum_{g=1}^G\sum_{h=1}^H N_{gh}$.

The way in which the regressors are generated inevitably affects the
finite\tkk-sample properties of every cluster-robust test statistic.
The differences between asymptotic and finite\tkk-sample distributions
arise mainly from the discrepancies between the disturbance vector
$\biu$ and the residual vector $\hat\biu = \biM_\biX\biu$. We use
\eqref{facDGP} to generate the regressor matrix $\biZ$ in \eqref{TWFE}
as well as the disturbances. In most experiments, we set
$\rho_g^x=\rho_h^x=0.2$ for the regressors and $\rho_g=\rho_h=0.1$ for
the disturbances. We use these values because, in practice, regressors
often display more intra-cluster correlation than residuals. For this
base case, we deliberately avoid situations, to be discussed below, in
which the amount of intra-cluster correlation is very small.

Throughout, when a standard error was either non-positive or undefined
for a particular replication, which can happen for CV$_{\tn1}^{(3)}$
or CV$_{\tn3}^{(3)}$, we counted it as a rejection.

\subsection{Cluster Size Variation}
\label{subsec:clustsize}

\begin{figure}[tb]
\begin{center}
\caption{Rejection frequencies as functions of how cluster sizes vary}
\label{fig:A}
\includegraphics[width=0.96\textwidth]{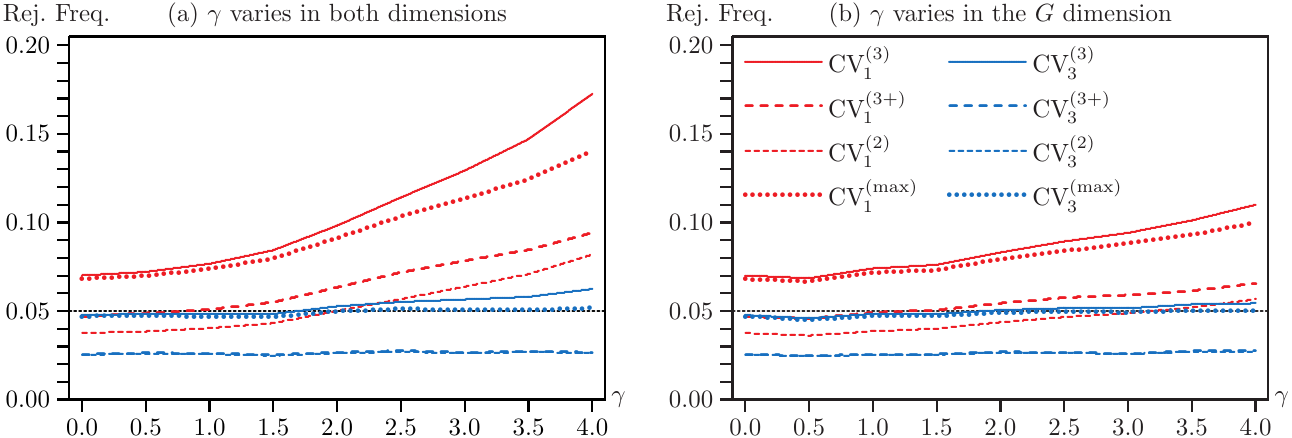}
\end{center}
{\footnotesize \textbf{Notes:} There are $N=9,\tn000$ observations,
with $G=15$, $H=12$, and $I=180$. The regressors and disturbances are
generated using \eqref{facDGP}, with $\rho_g^x=\rho_h^x=0.2$ for the
regressors and $\rho_g=\rho_h=0.1$ for the disturbances. The
regressand is generated using \eqref{TWFE} with all coefficients equal
to~0. The vertical axis shows rejection frequencies for $t$-tests at
the .05 level based on the $t(\min\{G,H\} -1)$ distribution. In 
Panel~(a), $\gamma$ is the same in both dimensions. In Panel~(b), the 
value of $\gamma$ is 0 for the $H$ dimension and varies for the $G$ 
dimension. The values of $p$ and $k$ are 10 and~36. There are 
100,\tk000 replications.}
\end{figure}

The first set of experiments focuses on cluster size variation,
determined by the parameter $\gamma$ in~\eqref{gameq}. \Cref{fig:A}
shows rejection frequencies for eight different $t$-tests from 18
experiments with $G=15$, $H=12$, and $N=9,\tn000$. In Panel~(a), the
value of $\gamma$ is varied simultaneously from 0.0 to 4.0 in both
dimensions. In Panel~(b), $\gamma=0$ for the $H$ dimension, and
$\gamma$ varies from 0.0 to 4.0 for the $G$ dimension. In these
experiments, the number of regressors is $p=10$. All tests would have
performed better if this had been a smaller number. The effects of
varying $p$ will be investigated below.

In Panel~(a), when all clusters are the same size (the leftmost point
on the horizontal axes), $t$-tests based on the classic
CV$_{\tn1}^{(3)}$ variance matrix estimator over-reject noticeably, as
do those based on CV$_{\tn1}^{(\max)}$. Rejection frequencies are 
considerably lower for CV$_{\tn1}^{(3+)}$, and lower still for
CV$_{\tn1}^{(2)}$. In contrast, $t$-tests based on the
CV$_{\tn3}^{(3)}$ and CV$_{\tn3}^{(\max)}$ estimators are very close
to nominal size, while those based on CV$_{\tn3}^{(3+)}$ and
CV$_{\tn3}^{(2)}$ under-reject substantially. As the value of~$\gamma$
increases, all the CV$_{\tn1}$ rejection frequencies rise sharply,
while those for the CV$_{\tn3}$ tests hardly change. Using the
\mbox{max-se} procedure has almost no effect when cluster sizes vary
little, but it modestly reduces rejection frequencies for CV$_{\tn1}$
tests when they vary a lot.

In Panel~(b), the overall patterns are similar. However, as predicted
in \Cref{subsec:whyjack}, rejection frequencies increase less rapidly
when $\gamma$ just increases in the $G$ dimension than when it
increases in both dimensions. In both panels, as must be the case,
$t$-tests based on two\tkk-term variance estimators always reject less
often than $t$-tests based on three\tkk-term ones. This is a good 
thing for the CV$_{\tn1}$ tests, but not for the CV$_{\tn3}$ ones.

One possibly surprising feature of \Cref{fig:A} is how much the 3+
tests based on the eigen-decomposition differ from the ordinary
three\tkk-term tests. This happens because, with 36 coefficients to
estimate (26 of them fixed effects), the three\tkk-term variance
matrices are not positive definite, leading to negative eigenvalues in
many cases. Whether or not the 3+ tests differ substantially from the
ordinary three\tkk-term tests depends on how the coefficient being
tested loads on the problematic eigen-directions. For this design, the
loading seems to be quite large. For CV$_{\tn1}$, the 3+ variant
performs better than the usual three\tkk-term test, but for
CV$_{\tn3}$, it performs worse, under-rejecting about as much as the
two\tkk-term test.

Except for quite small values of $\gamma$, the intersections in these
experiments vary greatly in size. For example, when both values of
$\gamma$ equal~2, which is the base case for many of our subsequent
experiments, the smallest intersection contains 6 observations, and
the largest contains~253. The sizes of the intersections vary much
more than those of the $G$ clusters, which range from 223 to~1443, or
the $H$ clusters, which range from 282 to~1769. Although these numbers
depend on the way in which we generate cluster sizes, it is inevitable
that, when the cluster sizes vary in both dimensions, the sizes of the
intersections vary more dramatically.

\Cref{fig:A} does not report results for the mixed variance estimator
\eqref{31jack}, or its \mbox{max-se} or 3+ variants, because the
results are very similar to the corresponding cluster-jackknife 
variants. For smaller values of~$\gamma$, the differences are 
negligible. The largest differences occur in Panel~(a) when
$\gamma=4$. In that case, the test based on $\hat\biV_3^{(3)}$ rejects
6.12\% of the time, while the one based on $\hat\biV_{3,1}^{(3)}$
rejects~5.55\%. The \mbox{max-se} tests differ much less, rejecting
5.35\% and 5.08\%, respectively, and the 3+ tests barely differ. In
most of our simulations, except when there were a great many empty
intersections (\Cref{subsec:empty}), the differences were much smaller
than in this case.

\subsection{Intra-Cluster Correlation}
\label{subsec:correlation}

\begin{figure}[tb]
\begin{center}
\caption{Rejection frequencies as functions of disturbance correlations}
\label{fig:B}
\includegraphics[width=0.96\textwidth]{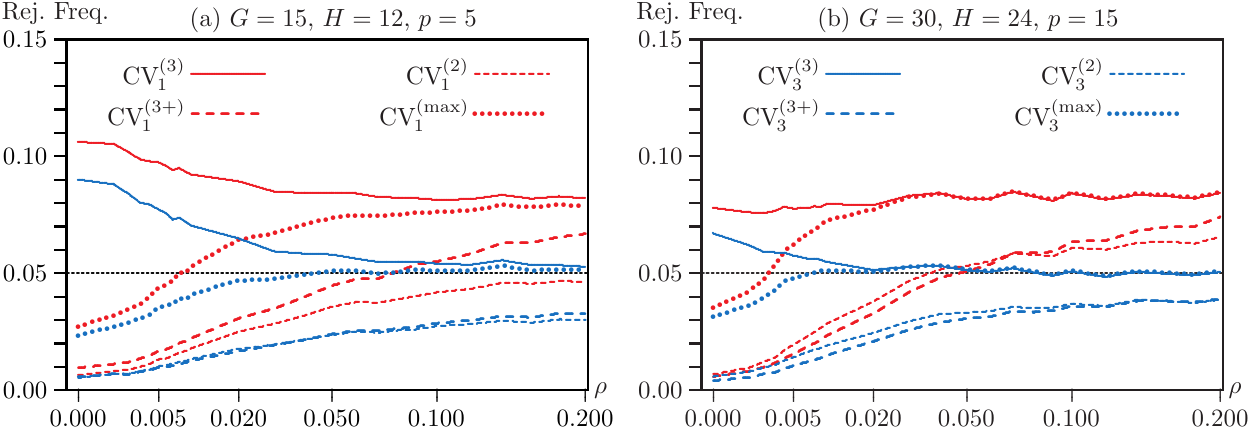}
\end{center}
{\footnotesize \textbf{Notes:} In Panel~(a), $N=9,\tn000$, with
$G=15$, $H=12$, $I=180$, and $p=5$. In Panel~(b), $N=36,\tn000$, with
$G=30$, $H=24$, $I=720$, and $p=15$. In both panels, $\gamma=2$ in
both dimensions. The $p$ regressors are generated using \eqref{facDGP}
with $\rho_g^x=\rho_h^x=0.2$. The disturbances are generated in the same
way, but with $\rho_g=\rho_h=\rho$, which varies from 0.000 to 0.010
by 0.001, from 0.010 to 0.100 by 0.010, and from 0.120 to 0.200 by
0.020. The regressand is generated using \eqref{TWFE} with all
coefficients equal to~0. The vertical axis shows rejection frequencies
for $t$-tests at the .05 level based on the $t(\min \{ G,H \} -1)$ 
distribution. The horizontal axis shows $\rho$, which is graphed on a
square\tkk-root scale. There are 100,\tk000 replications.}
\end{figure}

As \citet{MNW_2021} shows, test statistics based on the two\tkk-term
variance estimator are asymptotically too small whenever the scores
are asymptotically uncorrelated beyond the intersection level. This
suggests that they are likely to under-reject severely when the amount
of intra-cluster correlation in either the disturbances or regressors
is very small. In \Cref{fig:B}, we vary both values of $\rho$ for the
disturbances from 0.000 to~0.200. For clarity, the horizontal axis
uses a square root transformation. The numbers of clusters,
observations, and regressors are larger in Panel~(b) than in
Panel~(a); see the notes to the figure.

Several results stand out in \Cref{fig:B}. For small values of~$\rho$,
the rejection frequencies of the three\tkk-term tests are much higher
than those of the corresponding \mbox{max-se} tests. This is
particularly true for the CV$_{\tn1}$ tests. For the smallest values
of~$\rho$, the two\tkk-term tests under-reject to an extreme extent,
as the theory in \citet{MNW_2021} predicts. Interestingly, so do the
eigen-decomposition tests. In fact, in both panels, the two-term and
3+ tests perform very similarly. In both panels and for all values 
of~$\rho$, the CV$_{\tn3}$-based tests reject less than the 
corresponding CV$_{\tn1}$-based tests. Except for the smallest values 
of~$\rho$, the CV$_{\tn3}^{(3)}$ and CV$_{\tn3}^{(\max)}$ tests are
very similar in Panel~(a) and identical in Panel~(b), and they perform
very well.

For the smallest values of $\rho$ in these experiments, there were a
number of replications for which the three\tkk-term variance of
$\hat\beta_1$ was negative. This happened more often for $G=15$ than
for $G=30$, and more often for CV$_{\tn1}$ than for CV$_{\tn3}$. Since
we could not calculate the $t$-statistic for these replications, we
classified them as rejections. In the most extreme case, when
$\rho=0.000$ for $G=15$ ($G=30$), this happened 2.60\% (0.24\%) of the
time for CV$_{\tn1}^{(3)}$ and 2.1\% (0.16\%) for~CV$_{\tn3}^{(3)}$.
These numbers declined sharply as the value of $\rho$ increased.

\begin{figure}[tb]
\begin{center}
\caption{Rejection frequencies as functions of regressor correlations}
\label{fig:BB}
\includegraphics[width=0.96\textwidth]{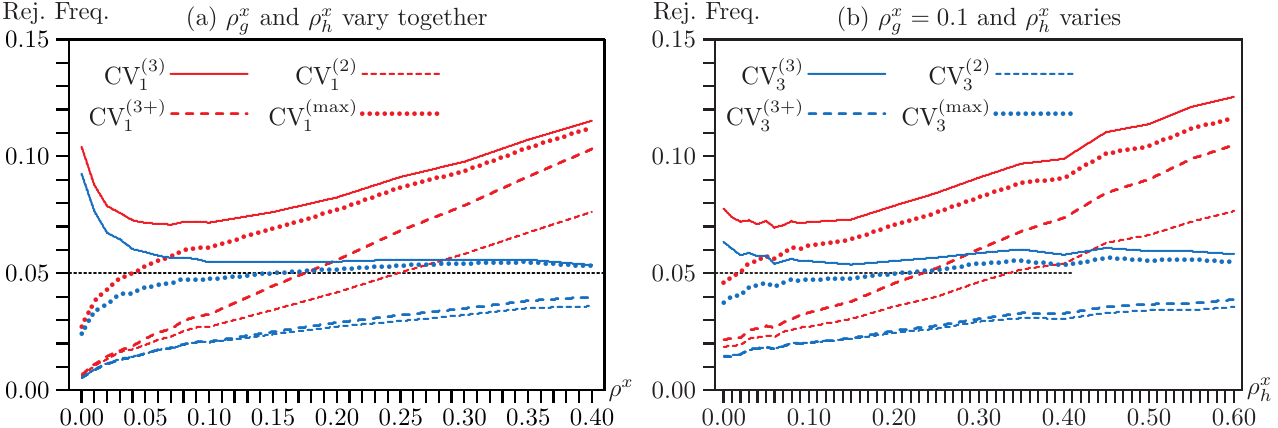}
\end{center}
{\footnotesize
\textbf{Notes:} In both panels, $N=9,\tn000$, with $G=15$, $H=12$,
$I=180$, $p=5$, and $\gamma=2$ in both dimensions. The disturbances
are generated using \eqref{facDGP} with $\rho_g=\rho_h=0.1$. The
regressors are also generated using \eqref{facDGP}, but the $\rho^x$
parameters vary. In Panel~(a), they both vary together between 0.00
and~0.30. In Panel~(b), $\rho_g^x=0.10$, and $\rho_h^x$ varies between
0.00 and~0.40. The regressand is generated using \eqref{TWFE} with all
coefficients equal to~0. The vertical axis shows rejection frequencies
for $t$-tests at the .05 level based on the $t(\min \{ G,H \} -1)$
distribution. There are 100,\tk000 replications.}
\end{figure}

It is not only the correlations of the disturbances that matter. In
\Cref{fig:BB}, we vary the correlations of the regressors, either in
both dimensions, in Panel~(a), or just in the $H$ dimension, in
Panel~(b).  Although they may seem small, the largest values of the
$\rho^x$ parameters here are not far short of the largest possible
values; see the discussion below~\eqref{facDGP}. The horizontal axis
does not use a square-root scale as \Cref{fig:B} did, because the
dependence on $\rho^x$ for small values is not as extreme as the
dependence on $\rho$ in that figure.

It is clear from \Cref{fig:BB} that the way in which the regressors
are distributed can have substantial effects on rejection frequencies.
Every test except CV$_{\tn1}^{(3)}$ and CV$_{\tn3}^{(3)}$ can either
over-reject or under-reject, depending on the values of the two
$\rho^x$ parameters. The two three\tkk-term tests always over-reject,
although only very slightly for CV$_{\tn3}^{(3)}$ in Panel~(a) for
$\rho^x>0.05$. The most reliable tests are the ones based on 
CV$_{\tn3}^{(3)}$ and, especially, CV$_{\tn3}^{(\max)}$. This is 
particularly the case for larger values of the $\rho^x$ parameters, 
where all the CV$_{\tn1}$-based tests over-reject substantially.
Panels (a) and (b) are quite similar when both $\rho^x$ parameters, or
just $\rho^x_h$, are large, but the two panels differ substantially
when the intra-cluster correlations are small.

\subsection{Number of Regressors}
\label{subsec:numregs}

\begin{figure}[tb]
\begin{center}
\caption{Rejection frequencies as functions of number of regressors}
\label{fig:C}
\includegraphics[width=0.96\textwidth]{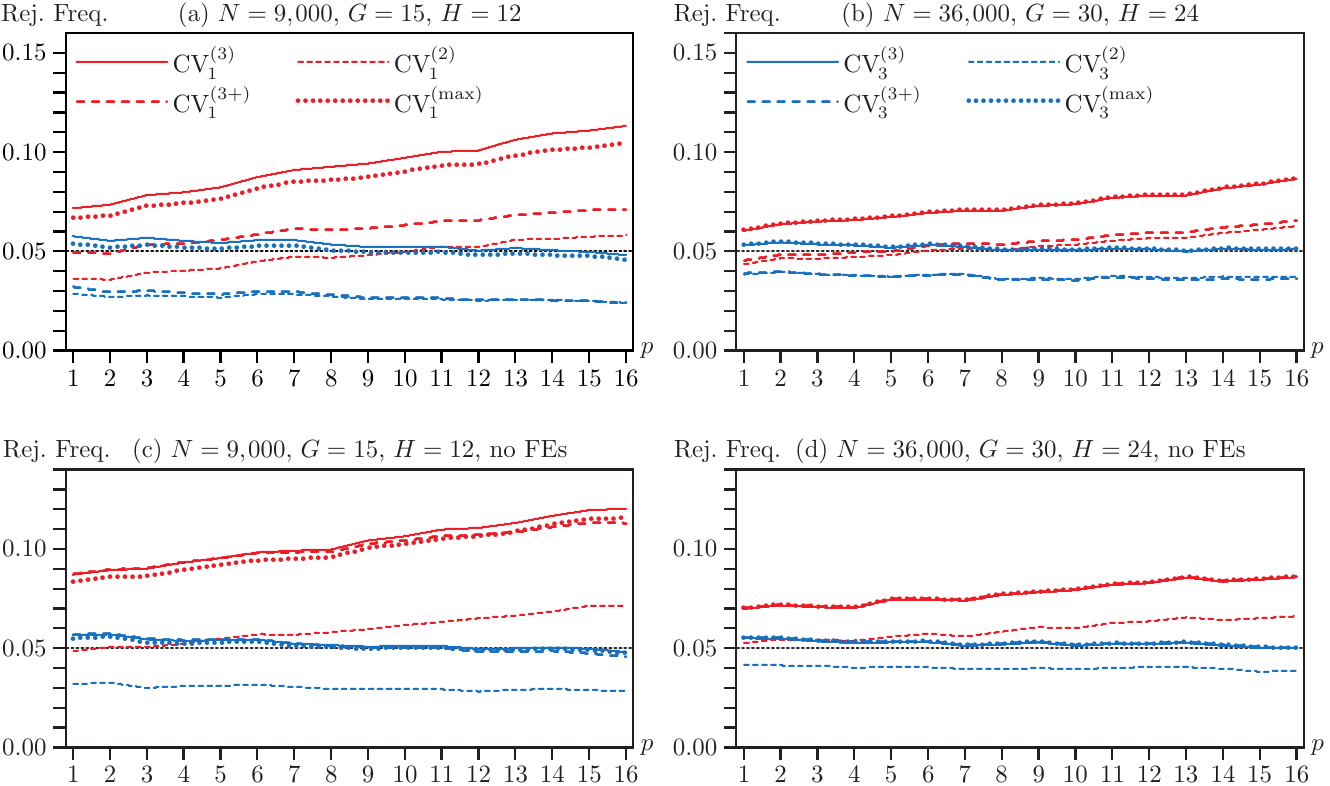}
\end{center}
{\footnotesize
\textbf{Notes:} There are $9,\tn000$ observations in Panels~(a)
and~(c) and $36,\tn000$ in Panels~(b) and~(d). The number of ordinary
regressors ($p$) varies from 1 to~16. They are generated using
\eqref{facDGP} with $\gamma=2$ and $\rho_g^x=\rho_h^x=0.2$. The
disturbances are generated in the same way, but with
$\rho_g=\rho_h=0.1$. In Panels~(a) and~(b), the regressand is
generated using \eqref{TWFE} with all coefficients equal to~0. In
Panels~(c) and~(d), all the fixed effects are replaced by a constant
term. The vertical axis shows rejection frequencies for $t$-tests at
the .05 level based on the $t(\min \{ G,H \} -1)$ distribution. There
are 100,\tk000 replications.}
\end{figure}

The number of regressors inevitably matters. This was analyzed in the
context of heteroskedasticity by \citet*{CJN_2018}. In \Cref{fig:C}, 
$p$ varies from 1 to~16. In Panels~(a) and~(c), $G=15$, $H=12$, and 
$N=9,\tn000$. In Panels~(b) and~(d), $G=30$, $H=24$, and 
$N=36,\tn000$. As $p$ increases, the rejection rates for the
CV$_{\tn1}$ tests increase, but those for the CV$_{\tn3}$ tests
decrease slightly. In all panels, the \mbox{max-se} tests perform
nearly the same as the three\tkk-term tests. Throughout \Cref{fig:C},
the CV$_{\tn3}^{(3)}$ and CV$_{\tn3}^{(\max)}$ tests perform very
well. 

In the lower two panels, the 26 or 53 fixed effects are replaced by a
constant term. Rejection frequencies for the CV$_{\tn1}$ tests still
increase with~$p$, but more slowly, while those for the CV$_{\tn3}$
tests still decrease, at about the same slow rate. In these two
panels, the 3+ tests are nearly identical to the ordinary three\tkk-term
tests. These are the only experiments in which we omit the fixed
effects. Their presence evidently has a large impact on the
performance of the 3+ tests but a fairly modest effect on that of the
other tests.

\Cref{fig:C} suggests that rejection frequencies for all the
CV$_{\tn1}$ tests increase fairly rapidly with~$p$, the number of
regressors that are not fixed effects, while those for all the
CV$_{\tn3}$ tests decrease quite slowly. We conjecture that this is
happening because all the regressors are correlated within clusters in
one or both dimensions. Thus, as the number of regressors increases,
more and more of the intra-cluster correlation in the disturbances is
explained by the regressors, so that less of it remains in the
residuals.

\begin{figure}[tb]
\begin{center}
\caption{Rejection frequencies as functions of number of extra
binary regressors}
\label{fig:G}
\includegraphics[width=0.96\textwidth]{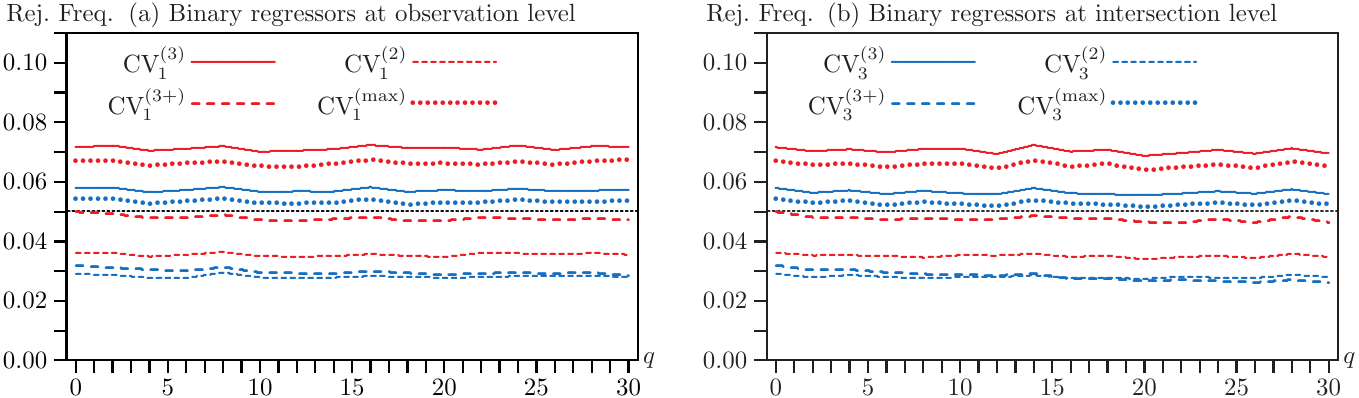}
\end{center}
{\footnotesize \textbf{Notes:} In both panels, $N=9,\tn000$, $G=15$,
$H=12$, and $\gamma=2$ in both dimensions. There is one continuous
regressor, like the ones in \Cref{fig:A,fig:B,fig:BB,fig:C,fig:D}. In
addition, there are $q$ binary regressors, which equal~1 with
probability~0.5. These vary independently at the observation
level in Panel~(a) and at the intersection level in Panel~(b). The 
vertical axis shows rejection frequencies for $t$-tests at the .05 level
based on the $t(\min \{ G,H \} -1)$ distribution. There are 100,\tk000 
replications.}
\end{figure}

In order to investigate this conjecture, we modify the way in which we
generate the regressors. The first one (the test regressor) is
generated as before, but then we generate an additional~$q$ binary
regressors which equal~0 or~1 with probability~0.5. In one set of
simulations, they are completely independent across observations. In a
second set, they are generated at the intersection level, identical
within each intersection and independent across intersections.

\Cref{fig:G} shows rejection frequencies as a function of~$q$, which 
varies from~0 to~30. In both panels, the test based on
CV$_{\tn3}^{(\max)}$ performs best, over-rejecting slightly for all
values of~$q$. The tests based on CV$_{\tn3}^{(3)}$, CV$_{\tn1}^{(3+)}$,
and CV$_{\tn1}^{(2)}$ perform nearly as well, with the former two 
over-rejecting slightly and the latter under-rejecting slightly. The 
value of $q$ has very little effect on most of the tests.

The differences between \Cref{fig:C} and \Cref{fig:G} are striking. In
the former, all the regressors are correlated within both the $G$ and
$H$ clusters. We saw there that adding more regressors with this
property can substantially increase rejection frequencies for
CV$_{\tn1}$ tests and slightly decrease them for CV$_{\tn3}$ tests. 
In contrast, adding more regressors that are uncorrelated across
observations or across intersections has almost no effect on rejection
frequencies. Empirical applications of two\tkk-way clustering often
involve many controls. Whether or not these controls exhibit
substantial correlation in either dimension can evidently be
important.

\subsection{Number of Clusters}
\label{subsec:numclust}

\begin{figure}[tb]
\begin{center}
\caption{Rejection frequencies as functions of number of clusters}
\label{fig:D}
\includegraphics[width=0.96\textwidth]{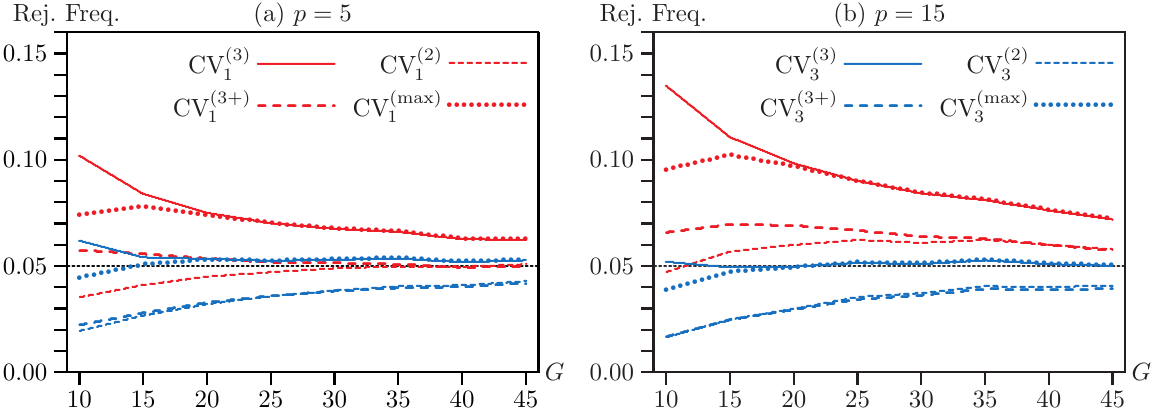}
\end{center}
{\footnotesize \textbf{Notes:} The value of $G$ varies from 10 to 45 
by~5, with $H=4G/5$ and $\gamma$ the same in both dimensions. There are
$50GH$ observations, so that $N$ varies from $4,\tn000$
to $81,\tn000$. The regressors are generated using \eqref{facDGP} with
$\rho_g^x=\rho_h^x=0.2$. The disturbances are generated in the same way,
but with $\rho_g=\rho_h=0.1$. The regressand is generated using
\eqref{TWFE} with all coefficients equal to~0. The vertical axis shows
rejection frequencies for $t$-tests at the .05 level based on the
$t(\min \{ G,H \} -1)$ distribution. There are 100,\tk000 replications.}
\end{figure}

All the results can be expected to improve as the number of clusters
increases in either or both dimensions. We saw this in \Cref{fig:C}.
Therefore, in \Cref{fig:D}, $G$ varies from 10 to 45 by~5, $H$ is
always equal to $4G/5$, and $N$ is proportional to~$GH$, so that the
sizes of the intersections are roughly constant. In Panel~(a), $p=5$.
Here, CV$_{\tn1}^{(3)}$ and CV$_{\tn1}^{(\max)}$, and to a lesser
extent CV$_{\tn3}^{(3+)}$, always over-reject, but they improve
steadily as $G$ (and~$H$) increase.  CV$_{\tn1}^{(\max)}$ over-rejects
less severely than CV$_{\tn1}^{(3)}$ for the smallest values of~$G$,
but the former is almost indistinguishable from the latter for
$G\ge15$. In contrast, CV$_{\tn3}^{(3)}$ always works almost
perfectly, with CV$_{\tn3}^{(\max)}$ yielding virtually identical
results for $G\ge15$.  By what seems to be coincidence,
CV$_{\tn1}^{(2)}$ also works well.

In Panel~(b) of \Cref{fig:D}, $p$ is increased to~15. The
CV$_{\tn1}$-based tests now over-reject much more severely, but tests
based on CV$_{\tn3}^{(3)}$ and CV$_{\tn3}^{(\max)}$ perform extremely
well. In contrast, tests based on CV$_{\tn3}^{(2)}$ and
CV$_{\tn3}^{(3+)}$ are almost identical and always under-reject. 
Clearly, omitting the intersection term or using the
eigen-decomposition is helpful for CV$_{\tn1}$, because the
three\tkk-term tests over-reject, but harmful for CV$_{\tn3}$, because
the three\tkk-term cluster-jackknife tests are approximately sized 
correctly.

\subsection{Empty Intersections}
\label{subsec:empty}

Up to this point, the data for all of our experiments have been
generated in such a way that $I=GH$. In other words, there have been
no datasets with empty intersections. But empirical examples with
two\tkk-way clustering often involve empty intersections. In the next
set of experiments, we therefore change the DGP so that intersections
can be empty. The details are somewhat complicated and are therefore
omitted. There is a parameter that determines the fraction of the
intersection clusters, starting with the smallest ones, to be made
empty by reallocating their observations proportionally to other
clusters. We perform two sets of experiments. In both of them, the 15
clusters in the $G$ dimension are generated from \eqref{gameq}
with~$\gamma=2$. In the first set $H=10$, and in the second
set~$H=20$. The maximum observed number of empty intersections is 95
(out of~150) in the first set and 270 (out of~300) in the second set.

\begin{figure}[tb]
\begin{center}
\caption{Rejection frequencies as functions of fraction of empty
intersections}
\label{fig:F}
\includegraphics[width=1.00\textwidth]{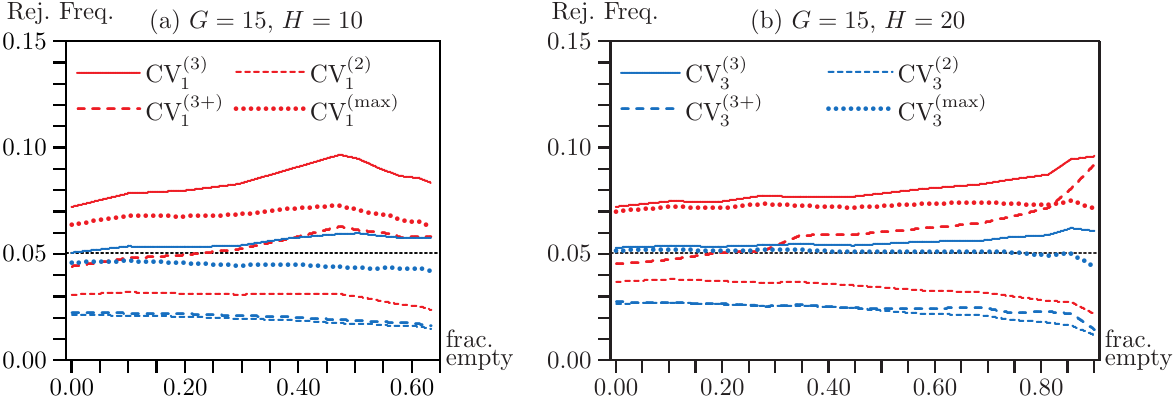}
\end{center}
{\footnotesize
\textbf{Notes:} In Panel~(a), $N=6,\tn000$. In Panel~(b), $N=12,\tn000$.
The disturbances are generated using \eqref{facDGP} with
$\rho_g=\rho_h=0.1$. There are 5 continuous regressors, which are
generated using \eqref{facDGP} with $\rho_g^x=\rho_h^x=0.2$, along
with 5 binary regressors, which vary at the intersection level and
equal 1 with probability~0.25. The fraction of empty intersections 
varies on the horizontal axis. The vertical axis shows rejection 
frequencies for $t$-tests at the .05 level based on the
$t(\min \{ G,H \} -1)$ distribution. There are 100,\tk000 replications.}
\end{figure}

\Cref{fig:F} shows rejection frequencies as functions of the fraction
of empty intersections. This fraction evidently matters, especially for 
the CV$_{\tn1}$-based tests, although not dramatically so in these 
experiments. As usual, $t$-tests based on CV$_{\tn3}^{(\max)}$ always 
perform best, and in fact they perform extremely well. Some of
the other tests perform quite poorly. As a rule, tests that
over-reject or under-reject when there are no empty intersections do
the same thing to a greater extent when there are many empty
intersections.

When the number of empty intersections is large, $I$ is not a great
deal larger than $G$ or~$H$, and the mixed variance estimators based
on \eqref{31jack} are no longer almost the same as the CV$_{\tn 3}$
ones. For the most extreme case, which is the rightmost point in 
Panel~(b) of \Cref{fig:F}, $I=30$ and $H=20$. In this case, the test 
based on $\hat\biV_3^{(3)}$ rejects 6.20\% of the time, while the one 
based on $\hat\biV_{3,1}^{(3)}$ rejects~4.82\%. For the \mbox{max-se} 
tests, the corresponding values are 4.53\% and~4.08\%.

\subsection{Test Power and Confidence Intervals}
\label{subsec:power}

\begin{figure}[tb]
\begin{center}
\caption{Power functions for eight tests}
\label{fig:7}
\includegraphics[width=0.50\textwidth]{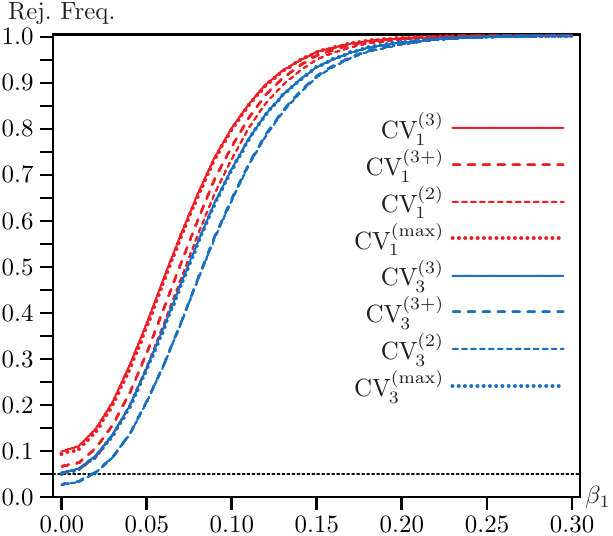}
\end{center}
{\footnotesize \textbf{Notes:} There are $9,\tn000$ observations,
with $G=15$, $H=12$, $p=10$, and $\gamma=2$ in both dimensions. The
regressors are generated using \eqref{facDGP} with
$\rho_g^x=\rho_h^x=0.2$. The disturbances are generated in the same
way, but with $\rho_g=\rho_h=0.1$. The regressand is generated using
\eqref{TWFE} with $\beta_1\ge0$ and all other coefficients equal to~0.
The vertical axis shows rejection frequencies for $t$-tests at the .05
level based on the $t(\min \{ G,H \} -1)$ distribution. There are
100,\tk000 replications.}
\end{figure}

Since some tests tend to over-reject and others tend to under-reject
under the null hypothesis, it is inevitable that the former will
appear to have more power than the latter. \Cref{fig:7} shows power
functions for all eight tests for a particular case. The functions
never cross, so there is nothing surprising here. For every
alternative, the ranking of the tests by power is identical to their
ranking by rejection frequencies under the null hypothesis. Thus the
fact that all of the CV$_{\tn1}$ tests appear to be more powerful than
any of the CV$_{\tn3}$ tests simply occurs because the former are
over-sized under the null. In this experiment, the power functions for
CV$_1^{\tk(\max)}$ and CV$_3^{\tk(\max)}$ are indistinguishable from
those for CV$_1^{\tk(3)}$ and CV$_3^{\tk(3)}$, respectively. All the
tests evidently reject with probability one when $\beta_1$ is 
sufficiently large.

We did not explicitly study coverage of confidence intervals, because
inverting tests that over-reject must yield intervals that
under-cover, and \textit{vice versa}. Similarly, inverting tests with
higher power will produce shorter intervals than inverting tests with
lower power. In these experiments, only the tests based on
CV$_3^{\tk(3)}$ and CV$_3^{\tk(\max)}$ have approximately correct
rejection frequencies, so only intervals based on them will have
approximately correct coverage. Although it would be possible to
obtain shorter intervals by using other standard errors, any such
intervals would be misleadingly short.

\section{Empirical Examples}
\label{sec:empirical}

In this section, we study two empirical examples. These involve
different types of two\tkk-way clustering and have clusters that
behave in different ways across the two clustering dimensions.

\subsection{The Tsetse Fly in African Development}
\label{subsec:tsetse}

\begin{table}[tp]
\caption{Empirical results for tsetse fly example}
\label{tab:tsetse}
\vspace*{-0.5em}
\begin{tabular*}{\textwidth}{@{\extracolsep{\fill}}lccccccccc}
\toprule
\multicolumn{2}{l}{Panel~A: Conventional CRVEs}
& \multicolumn{7}{c}{$P$~values} \\
\cmidrule{3-9}
Dependent variable& $\hat\beta$ & \textrm{HC$_1$} & \textrm{CV$_{\tn1}$-I}
& \textrm{CV$_{\tn1}$-G} & \textrm{CV$_{\tn1}$-H} 
& \textrm{CV$_{\tn1}^{(2)}$} & \textrm{CV$_{\tn1}^{(3)}$} 
& \textrm{CV$_{\tn1}^{(\max)}$}\\
\midrule
Large animals& $-0.2310$ & $0.0000$  &$0.0000$ &$0.0000$ &$0.0000$ 
&$0.0002$ &$0.0000$ &$0.0000$ \\
Intensive agriculture& $-0.0905$ & $0.0080$  &$0.0087$ &$0.0020$ &$0.0045$ 
&$0.0321$ &$0.0003$ &$0.0045$ \\
Plow use& $-0.0572$ & $0.0096$  &$0.0171$ &$0.0149$ &$0.0791$ 
&$0.1496$ &$0.0715$ &$0.0791$ \\
Female participation& $\phantom{-}0.2057$ & $0.0001$  &$0.0001$ &$0.0014$ &$0.0001$ 
&$0.0099$ &$0.0009$ &$0.0014$ \\
Log pop.\ density& $-0.7446$ & $0.0000$  &$0.0002$ &$0.0023$ &$0.0005$ 
&$0.0182$ &$0.0028$ &$0.0028$ \\
Indigenous slavery& $\phantom{-}0.1013$ & $0.0060$  &$0.0137$ &$0.0160$ &$0.0162$ 
&$0.0834$ &$0.0157$ &$0.0162$ \\
Centralization& $-0.0746$ & $0.0727$  &$0.0460$ &$0.0395$ &$0.0471$ 
&$0.1482$ &$0.0365$ &$0.0471$ \\
\midrule
\\[-10pt]
\multicolumn{2}{l}{Panel~B: Jackknife CRVEs} 
& \multicolumn{7}{c}{$P$~values} \\
\cmidrule{3-9}
Dependent variable& $\hat\beta$ & \textrm{HC$_3$} & \textrm{CV$_{\tn3}$-I}
& \textrm{CV$_{\tn3}$-G} & \textrm{CV$_{\tn3}$-H} 
& \textrm{CV$_{\tn3}^{(2)}$} & \textrm{CV$_{\tn3}^{(3)}$} 
& \textrm{CV$_{\tn3}^{(\max)}$}\\
\midrule
Large animals& $-0.2310$ & $0.0000$  &$0.0000$ &$0.0000$ &$0.0000$ 
&$0.0007$ &$0.0000$ &$0.0000$ \\
Intensive agriculture& $-0.0905$ & $0.0092$  &$0.0130$ &$0.0044$ &$0.0123$ 
&$0.0549$ &$0.0027$ &$0.0123$ \\
Plow use& $-0.0572$ & $0.0124$  &$0.0220$ &$0.0207$ &$0.1422$ 
&$0.2112$ &$0.1365$ &$0.1422$ \\
Female participation& $\phantom{-}0.2057$ & $0.0001$  &$0.0002$ &$0.0036$ &$0.0008$ 
&$0.0239$ &$0.0051$ &$0.0051$ \\
Log pop.\ density& $-0.7446$ & $0.0000$  &$0.0005$ &$0.0060$ &$0.0021$ 
&$0.0357$ &$0.0090$ &$0.0090$ \\
Indigenous slavery& $\phantom{-}0.1013$ & $0.0077$  &$0.0247$ &$0.0353$ &$0.0496$ 
&$0.1462$ &$0.0575$ &$0.0575$ \\
Centralization& $-0.0746$ & $0.0798$  &$0.0564$ &$0.0545$ &$0.0817$ 
&$0.1925$ &$0.0751$ &$0.0817$ \\
\midrule
\\[-10pt]
\multicolumn{2}{l}{Panel~C: Cluster diagnostics} \\[-15pt]
& \multicolumn{6}{c}{Coefficients of variation} \\
\cmidrule{2-7}
& \multicolumn{3}{c}{partial leverage} 
& \multicolumn{3}{c}{$\hat\beta^{(g)}$} & \multicolumn{2}{c}{$G^{*}$} \\
\cmidrule{2-4}\cmidrule{5-7}\cmidrule{8-9}
Dependent variable & culture & country & intersect & culture & country 
& intersect & culture & country \\
\midrule
Large animals & 1.2877 & 1.1571 & 1.7094 & 0.0323 & 0.0271 & 0.0173 & 23.18 & 16.25 \\
Intensive agriculture & 1.2875 & 1.1572 & 1.7095 & 0.0513 & 0.0591 & 0.0336 & 23.20 & 16.27 \\
Plow use & 1.2877 & 1.1571 & 1.7094 & 0.0643 & 0.1035 & 0.0365 & 23.18 & 16.25 \\
Female participation & 1.6073 & 1.4962 & 2.0302 & 0.0507 & 0.0444 & 0.0248 & 20.84 & 13.71 \\
Log pop.\ density & 1.3241 & 1.2277 & 1.8453 & 0.0540 & 0.0481 & 0.0258 & 20.88 & 15.51 \\
Indigenous slavery & 1.3307 & 1.2431 & 1.7912 & 0.0710 & 0.0774 & 0.0376 & 22.71 & 15.84 \\
Centralization & 1.3189 & 1.1751 & 1.7403 & 0.0780 & 0.0861 & 0.0442 & 22.76 & 16.20 \\
\bottomrule
\end{tabular*}
\vskip 6pt {\footnotesize \textbf{Notes:} Estimates correspond to
Table~1~(4) and Table~3~(8) from \citet{Alsan_2015}. Panels A and B
show coefficient estimates and $P$~values. CV-G is clustered by
cultural province, CV-H is clustered by country, and CV-I is clustered
by intersection. $P$~values for HC standard errors are based on the
$t(N-13)$ distribution. $P$~values for CV-G, CV-H, and CV-I are based
on the $t(G-1)$, $t(H-1)$, or \mbox{$t(I-1)$} distributions,
respectively. $P$~values for two\tkk-way clustering are based on the
$t(\min\{G,H\}-1)$ distribution. In all cases, the $P$~values based on
CV$_{\tn3}^{(3+)}$ are identical to those based on CV$_{\tn3}^{(3)}$
to the number of digits reported. Panel~C shows coefficients of
variation for partial leverage and omit-\tkk one\tkk-cluster
coefficients by both clustering dimensions and their intersection, as
well as the effective number of clusters, $G^*=G^*(0)$, computed by
\texttt{twowayjack}.}
\end{table}

In a fascinating paper, \citet{Alsan_2015} studies the effects of the
tsetse fly on African development. The key explanatory variable is the
``tsetse suitability index,'' or TSI, which measures the extent to
which climate (temperature and humidity) is suitable for the tsetse
fly to thrive. There are seven dependent variables, which measure
various aspects of economic and political development. Each of these
is regressed on the TSI, whose coefficient is denoted $\beta$, and on 
eleven other variables in the columns labeled ``(4)'' in Table~1 and
``(8)'' in Table~3 of \citet{Alsan_2015}. The former uses one\tkk-way 
clustering by ``cultural province'' and the latter uses two\tkk-way
clustering by cultural province and country. There are 44 countries
and either 43 or 44 cultural provinces, depending on the regressand.
Since the total number of observations varies between 315 and~485,
most clusters are quite small, and there are many empty intersections.
The number of non-empty intersections varies between 112 and~142.

In \Cref{tab:tsetse}, we report $P$~values based on sixteen different
standard error estimates, eight using conventional standard errors
(Panel~A) and eight using jackknife ones (Panel~B). Because the
ordinary three\tkk-term and eigen-decomposition three\tkk-term
standard errors are identical in all cases (to the number of digits
reported), we only report the former.

Even though the clusters are quite small (the largest is 63, which is 
for clustering by country when the dependent variable is the log of
population density), the way in which we cluster often makes a
substantial difference. Not clustering at all sometimes leads to
extremely small $P$~values, as does one\tkk-way clustering by
intersection. Clustering in two dimensions often, but not always,
leads to larger $P$~values than clustering in just one dimension. For
two\tkk-way clustering, the cluster-jackknife $P$~values are never
smaller than the conventional ones, and they are mostly considerably 
larger. We also calculated $P$~values based on the mixed three-term
estimator \eqref{31jack}, but we do not report them because, with so
many empty intersections, it is hard to justify using the mixed
estimator. They are very similar to the ones in Panel~B, although
slightly larger in all but one case.

Panel~C of \Cref{tab:tsetse} presents a number of the summary
statistics calculated by \texttt{twowayjack} for this example.
Specifically, it presents coefficients of variation for the partial
leverages of the TSI variable and for the $\hat\beta^{(g)}$ for
clustering by cultural province, country, and intersection. It also
displays the effective number of clusters $G^*=G^*(0)$ for the two
primary dimensions \citep{CSS_2017,MNW-influence}. These diagnostics
can help to explain why some of the $P$~values in Panels~A and~B
differ by more than others. The notable $P$~value differences between
CV$_{\tn3}^{(\max)}$ and CV$_{\tn1}^{(\max)}$ occur for `plow use,'
`indigenous slavery,' and `centralization.' For these three variables,
we see the largest coefficients of variation for the omit-\tkk
one\tkk-country estimates, and, to a slightly lesser extent, for the
omit-\tkk one\tkk-culture ones.

The results of \Cref{sec:simuls} suggest that CV$_{\tn 3}^{(\max)}$
yields the most reliable $P$~values. The CV$_{\tn 3}^{(\max)}$
$P$~value for TSI is less than 0.05 for four of the seven dependent
variables. In contrast, the CV$_{\tn1}$ $P$~values for one\tkk-way
clustering by cultural province used in \citet{Alsan_2015} are less
than 0.05 for all seven variables in Table~1~(4), and the ones for
two\tkk-way CV$_{\tn1}^{(3)}$ clustering are less than 0.05 for six of
them in Table~3~(8). Thus, although there is still a good deal of
evidence that the TSI matters for a variety of outcomes, the evidence
is not quite as strong as it originally seemed to be.

\subsection{Minimum Wages in Canada}
\label{subsec:minwage}

Our second example examines the relationship between minimum wages in
Canada and the log of hourly earnings. We focus attention on men
between 18 and 24 years of age who immigrated to Canada less than ten
years ago. Our sample contains 28,599 observations for the years 2008
to~2019. Except for a few federally-regulated industries, minimum
wages in Canada are set at the provincial level. They tend to change
infrequently, and they never go down. In fact, although our sample
contains observations for 1440 province-month pairs, the minimum wage
variable takes on only 63 unique values.

The equation we estimate is
\begin{equation}
\text{logearn}_{ipmt} = \alpha + \beta\, \text{logmw}_{\tn pmt} 
  + \gamma\, \text{bigcity}_{\tn ipmt} + \delta\,\text{age}_{ipmt} 
  + \text{year}_t + \text{month}_m + \text{prov}_{\tn p}
  + \epsilon_{ipmt},
\label{minwage}
\end{equation}
where logearn$_{ipmt}$ is the log of hourly earnings for individual~$i$ 
in province~$p$ in month~$m$ of year~$t$, logmw$_{\tn pmt}$ is the log 
of the minimum wage, bigcity$_{\tn ipmt}$ is a dummy for being in one
of nine large cities, age$_{ipmt}$ is a dummy for being 22 to~24, and 
the remaining regressors are year fixed effects, calendar month 
fixed effects, and province fixed effects. The total number of 
regressors, including the constant term, is~35.

This example is one for which reliable cluster-robust inference is
likely to be difficult. We cluster by year and province, but there are
only 12 years and 10 provinces. The year clusters are reasonably
homogeneous in size; they vary from 2051 to 2723 observations. But the
province clusters are very heterogeneous; they vary from 163 (P.E.I.)
to 6554 (Ontario). Although there are no empty intersections, the
smallest contains just 3 observations, and the largest contains~710.

\begin{table}[tp]
\caption{Empirical results for minimum wage example}
\label{tab:minwage}
\vspace*{-0.5em}
\begin{tabular*}{\textwidth}{@{\extracolsep{\fill}}lccccccccc}
\toprule
\multicolumn{9}{l}{Panel~A: Conventional CRVEs} \\
& \textrm{HC$_1$} & \textrm{CV$_{\tn1}$-I}
& \textrm{CV$_{\tn1}$-G} & \textrm{CV$_{\tn1}$-H} 
& \textrm{CV$_{\tn1}^{(2)}$} & \textrm{CV$_{\tn1}^{(3)}$} 
& \textrm{CV$_{\tn1}^{(3+)}$} & \textrm{CV$_{\tn1}^{(\max)}$}\\
\midrule
$P$~values & $0.0000$  &$0.0000$ &$0.0041$ &$0.0034$ 
&$0.0261$ &$0.0140$ &$0.0141$ &$0.0140$ \\
Placebo rej.\ freq. & $0.8947$  &$0.6301$ &$0.5958$ &$0.3018$ 
&$0.1454$ &$0.2431$ &$0.2431$ &$0.2319$ \\
\midrule
\\[-10pt]
\multicolumn{9}{l}{Panel~B: Jackknife CRVEs} \\
& \textrm{HC$_3$} & \textrm{CV$_{\tn3}$-I}
& \textrm{CV$_{\tn3}$-G} & \textrm{CV$_{\tn3}$-H} 
& \textrm{CV$_{\tn3}^{(2)}$} & \textrm{CV$_{\tn3}^{(3)}$} 
& \textrm{CV$_{\tn3}^{(3+)}$} & \textrm{CV$_{\tn3}^{(\max)}$}\\
\midrule
$P$~values & $0.0000$  &$0.0001$ &$0.0125$ &$0.0565$ 
&$0.1116$ &$0.0808$ &$0.0810$ &$0.0808$ \\
Placebo rej.\ freq.& $0.8947$  &$0.5725$ &$0.5432$ &$0.0896$ 
&$0.0254$ &$0.0649$ &$0.0649$ &$0.0572$ \\
\midrule
\\[-10pt]
\multicolumn{2}{l}{Panel~C: Cluster diagnostics} \\[-15pt]
& \multicolumn{6}{c}{Coefficients of variation} \\
\cmidrule{2-7}
& \multicolumn{3}{c}{partial leverage} 
& \multicolumn{3}{c}{$\hat\beta^{(g)}$} & \multicolumn{2}{c}{$G^{*}$} \\
\cmidrule{2-4}\cmidrule{5-7}\cmidrule{8-9}
& year & province & intersect & year & province 
& intersect & year & province \\
\midrule
Log earnings & 0.0607 & 1.1909 & 1.1794 & 0.1061 & 0.1577 & 0.0232 &
6.51 & 4.49 \\
\bottomrule
\end{tabular*}
\vskip 6pt {\footnotesize \textbf{Notes:} There are 28,599
observations, 12 year ($G$) clusters, 10 province ($H$) clusters, and
120 intersection ($I$) clusters. The coefficient estimate for the log
minimum wage is~$\hat\beta=0.2934$. The first row in each of Panels~A
and~B reports $P$~values using HC standard errors based on the
$t(28,\tn564)$ distribution, $P$~values for one\tkk-way clustering
based on the $t(11)$, $t(9)$, or $t(119)$ distributions, as
appropriate, and $P$~values for two\tkk-way clustering based on the
$t(9)$ distribution. The second row reports rejection frequencies for
100,\tk000 placebo regressions. Panel~C reports coefficients of variation
for partial leverage and omit-\tkk one\tkk-cluster coefficients by
both clustering dimensions and their intersection, as well as the
effective number of clusters, $G^*=G^*(0)$, computed by
\texttt{twowayjack}.}
\end{table}

\Cref{tab:minwage} contains three panels. Panel~C presents some
cluster diagnostics, calculated using \texttt{twowayjack}. The
coefficients of variation are quite revealing. For partial leverage,
there is considerable variation across provinces and intersections,
but very little across years. For the $\hat\beta^{(g)}$, there is
modest variation when leaving out a province or a year, but very
little when leaving out an intersection cluster.

These features of the sample suggest that many methods, perhaps all
methods, will not yield reliable inferences. In order to investigate
this conjecture, we employ placebo\tkk-regression simulations as advocated 
by \citet[Section~3.5]{MNW-guide}. These are similar in spirit to the 
``placebo laws'' simulations of \citet*{BDM_2004}. For each of 100,\tk000
simulations and each province, we generate a sequence of values of a
placebo regressor that resembles the actual minimum wage sequences:
The value tends to stay constant for a while and then rise by a random
amount from time to time in a fashion that is correlated across
provinces. This placebo regressor is then added to
regression~\eqref{minwage}, and we calculate sixteen $P$~values for
its coefficient based on all sixteen standard errors used for the
actual regression. If regression~\eqref{minwage} is correctly
specified and any particular way of obtaining $P$~values is valid for
our sample, then the fraction of the time that the placebo\tkk-regression
$P$~value is less than 0.05 should be very close to~0.05, subject to
experimental error.

Panels~A and~B of \Cref{tab:minwage} show both the actual $P$~values
and rejection frequencies for the placebo regressions for all sixteen
methods. The conventional $P$~values in Panel~A imply that the minimum
wage is significant at the 0.01 level for all the one\tkk-way
clustering methods and at the 0.02 level for all the two\tkk-way
methods. However, the placebo\tkk-regression rejection frequencies vary
from 15\% to~89\%, suggesting that none of the conventional $P$~values
should be believed.

In contrast, the jackknife $P$~values in Panel~B are greater than 0.05
for one\tkk-way clustering by province and for the two\tkk-way
clustering methods. The placebo\tkk-regression rejection frequencies for
the one\tkk-way methods vary between~9\% and~89\%, suggesting that
they should not be trusted. For the two\tkk-term two\tkk-way estimator
CV$_{\tn3}^{(2)}$, the placebo rejection frequency is just~2.5\%.
This is in line with existing theory (\Cref{sec:concepts}) and some
of our simulations (e.g.\ \Cref{fig:B}), which both show that 
CV$_{\tn3}^{(2)}$ can under-reject. For the three\tkk-term
cluster-jackknife estimators, the placebo rejection frequencies are
between 5.7\% and~6.5\%, which is remarkably good in view of the small
numbers of clusters and the cluster diagnostics. For these methods,
the $P$~value is~0.081.

Because the number of clusters in each of the primary dimensions is
small, and cluster sizes vary greatly in one of them, it seems
plausible that even the three-term CV$_{\tn3}$ standard errors are too
small. The modest over-rejection rates for the placebo regressions in
the second row of the last three columns of Panel~B support this
conjecture. Thus it could makes sense to use the mixed three-term
estimator \eqref{31jack} in this case. The placebo regression
rejection rates for CV$_{3,1}^{(3)}$ and CV$_{3,1}^{(\rm max)}$ are
0.0506 and 0.0479, respectively. Like that of CV$_{3}^{(\rm max)}$,
these are extraordinarily close to~0.05. The $P$~values for both
CV$_{3,1}^{(3)}$ and CV$_{3,1}^{(\rm max)}$ are 0.0954, which is also
close to that of CV$_{3}^{(\rm max)}$ at~0.0808.

We conclude that, even though all the conventional one\tkk-way and
two\tkk-way standard errors yield strongly significant results (in the
first row of Panel~A), all the two\tkk-way jackknife standard errors
(in the first row of Panel~B and in the previous paragraph) yield
results that are not even close to significant at the 0.05 level. Thus,
we cannot be confident that the minimum wage positively affects hourly
earnings based on the evidence from this sample.

\section{Conclusions}
\label{sec:conclusions}

It is common to assume that the disturbances in linear regression
models are clustered in two dimensions. Unless the regressor(s) of
interest are uncorrelated in one or both dimensions, it is therefore
necessary to employ a cluster-robust variance estimator that allows
for two\tkk-way clustering. Unfortunately the most widely-used 
cluster-robust variance matrix estimator (CRVE), CV$_{\tn1}^{(3)}$, due 
to \citet{CGM_2011}, is not guaranteed to be positive definite.
Inferences based on it are known to be seriously unreliable in finite
samples \citep{MNW_2021}.

In \Cref{sec:concepts}, we discuss several ways to avoid, or at least
ameliorate, the problem of undefined standard errors when a CRVE is
not positive definite. Most importantly, we propose a new and simple
solution to this problem. For tests of a single restriction, it just
involves using whichever of three standard errors is the largest. Two
of these are based on one\tkk-way clustering in each of the two
dimensions, and the third is a three\tkk-term two\tkk-way standard
error. Asymptotically, the latter should always be the largest of the
three when there really is two\tkk-way clustering, but it may not be
the largest (and may indeed not be defined) in finite samples. In many
cases, our so\tkk-called \mbox{max-se} procedure yields results
identical to those from the corresponding three\tkk-term two\tkk-way
CRVE, but it can yield substantially lower (and more accurate)
rejection frequencies in some cases.

The second, and in our view more widely applicable, contribution of
the paper is to propose and study two\tkk-way cluster-jackknife CRVEs.
Recent work on the cluster-jackknife, or CV$_{\tn3}$, CRVE for
one\tkk-way clustering
\citep{Hansen-jack,Hansen_2025,MNW-bootknife,MNW-influence} suggests
that it can perform much better in finite samples than the usual
CV$_{\tn1}$ CRVE. It therefore seems attractive to extend it to the
two\tkk-way case. This is remarkably simple. We just need to perform
three sets of cluster-jackknife calculations, one for each of the two
dimensions, and then a third one for their intersections. In many
cases, this is straightforward, although cluster fixed effects do
raise some computational issues (\Cref{sec:twojack}), and the
calculations can be costly when the number of intersections is large,
especially when there are cluster fixed effects. We provide a
\texttt{Stata} package called \texttt{twowayjack} that implements our
methods and also calculates some cluster diagnostics
\citep{MNW-influence}; see \Cref{app:software}.

In \Cref{sec:simuls}, we study rejection frequencies for $t$-tests
based on eight different cluster-robust standard errors. Four of them
are of the usual CV$_{\tn1}$ type, and the other four are of the
CV$_{\tn3}$ type. In most cases, tests based on the CV$_{\tn3}$
\mbox{max-se} standard error yield the most reliable inferences. Even
when they do not, they only perform slightly worse than whatever
procedure(s) perform better, and they are usually much more reliable
than all of the CV$_{\tn1}$-based tests.

Because most of our simulations involve two\tkk-way cluster fixed
effects, three\tkk-term variance matrices based on either CV$_{\tn1}$
or CV$_{\tn3}$ tend not to be positive definite, so the versions that
use an eigen-decomposition (\Cref{sec:concepts}) can differ greatly
from the versions that do not. This always reduces rejection
frequencies, which is a good thing for CV$_{\tn1}$ tests but usually a
bad thing for CV$_{\tn3}$ tests. Tests based on two\tkk-term variance
matrices usually reject even less frequently than tests based on
three\tkk-term variance matrices with the eigen-decomposition. Thus we
do not recommend using tests based on either CV$_{\tn3}^{(2)}$ or
CV$_{\tn3}^{(3+)}$.

Our simulations show that precisely how the data are generated can
have a large effect on finite\tkk-sample performance. All the tests
are most likely to perform poorly when the number of clusters in
either dimension is small, cluster sizes vary greatly, there are many
empty intersections, the number of regressors that are clustered in
one or both dimensions is large, or either the disturbances or the
regressor(s) of interest are only weakly correlated in both
dimensions. In many of these cases, alternative test statistics tend
to perform quite differently. For example, weak intra-cluster
correlation is particularly problematic for the CV${\tn1}$-based,
two-term, and 3+ variants. The CV$_{\tn3}^{(\max)}$ tests generally
stay close to their nominal size except in the most extreme
weak-correlation designs or when combined with other adverse features.

In practice, it can often be illuminating to employ placebo regression
simulations, as in \Cref{subsec:minwage}. These will show how well
alternative tests perform for the particular model and dataset under
study. It is probably safe to rely on CV$_{\tn3}^{\tk(\max)}$-based
tests if they perform well in these simulations, or perhaps on some
other tests if they perform better.

\appendix 
\numberwithin{equation}{section} 
\numberwithin{figure}{section} 
\numberwithin{table}{section} 

\makeatletter 
\def\@seccntformat#1{\@ifundefined{#1@cntformat}
	{\csname the#1\endcsname\quad}
	{\csname #1@cntformat\endcsname}}
\newcommand\section@cntformat{}
\makeatother

\section{Appendix A: Proof of \Cref{thm:cons}}
\label{app:proof}

Recall that $J \in \{ G, H, I \}$, where $j$ denotes the 
corresponding lower-case letter. In the intersection dimension
($J=I$), the summation $\sum_{j=1}^J \biZ_j$ should be interpreted as 
$\sum_{g=1}^G\sum_{h=1}^H \biZ_{gh}$ for any $\biZ$, where $I$ denotes
the number of non-empty intersections, which may be smaller than~$GH$.

We first present a useful lemma. The proof is almost identical to that 
of Lemma~A.2 in \citet{DMN_2019}, and is therefore omitted.

\begin{lemma}
\label{lem:orders}
Under \Cref{assn:yap},
\begin{equation*}
\max_{1\leq j \leq J} N_j^{-1} \Vert \bis_j \Vert = O_P (1) \text{ and }
\max_{1\leq j \leq J} N_j^{-1} \Vert \biX_j^\top\biX_j \Vert = O_P(1).
\end{equation*}
\end{lemma}

Let us define $\biW_j^\top = (\biW_{1,j} , \biW_{2,j}) = (\biX_j^\top
\biX_j , \bis_j )$ using an array notation, and let $\bar\biW
=N^{-1}\sum_{j=1}^J \biW_j$. Under \Cref{assn:yap}, we can apply the
law of large numbers and consistent variance estimation arguments in 
\citet[e.g., Proposition~2 and Lemma~8]{yap_2025}, so that
\begin{align}
\bar\biW - \bmu &\Pto \bzero ,\label{lln1} \\
\sum_{g=1}^G \bis_g \bis_g^\top 
+\sum_{h=1}^H \bis_h \bis_h^\top 
- \sum_{g=1}^G \sum_{h=1}^H \bis_{gh} \bis_{gh}^\top 
&= \bSigma \big(\bfI_k + o_P(1)\big), \label{lln4} 
\end{align}
where $\bmu^\top = ( \biS_{xx}, \bzero )$, and 
$\biS_{xx}=N^{-1}\E(\biX^\top\biX)$ is invertible by \Cref{assn:yap}(e).

With a slight abuse of notation, let $\bar\biW^{(j)}=N^{-1}
\sum_{m=1,m\neq j}^J \biW_m$. From \Cref{lem:orders}, we find 
$\max_{1\leq j \leq J} \Vert \bar\biW-\bar\biW^{(j)} \Vert 
= N^{-1} \max_{1\leq j \leq J} \Vert \biW_j \Vert 
= \max_{1\leq j \leq J} O_P(N^{-1}N_j) \Pto 0$, where the convergence 
is a consequence of \Cref{assn:yap}(b). It then follows from 
\eqref{lln1} that 
\begin{equation}
\label{lln5}
\bar\biW^{(j)} - \bmu \Pto \bzero \text{ for all }j=1,\ldots ,J.
\end{equation}
With this notation, $\hat\bbeta-\bbeta_0=f(\bar\biW)$ and 
$\hat\bbeta^{(j)} - \bbeta_0 = f(\bar\biW^{(j)})$ for the mapping 
$f(\biZ)=\biZ_1^{-1}\biZ_2$, with~$\biZ^\top = (\biZ_1,\biZ_2)$, where 
the $\hat\bbeta^{(j)}$ exist due to \Cref{assn:rank}. By the mean value 
theorem,
\begin{equation*}
\hat\bbeta^{(j)} - \hat\bbeta 
= (\hat\bbeta^{(j)} - \bbeta_0 ) - (\hat\bbeta - \bbeta_0 )
= f (\bar\biW^{(j)} ) - f (\bar\biW ) 
= \nabla f ( \bxi_j )^\top (\bar\biW^{(j)} - \bar\biW ) ,
\end{equation*}
where $\nabla f$ denotes the gradient of~$f$, and $\bxi_j$ is a point 
on the line segment between $\bar\biW^{(j)}$ and~$\bar\biW\!$. Using 
also $\bar\biW^{(j)} - \bar\biW = -N^{-1}\biW_j$, we then find that
\begin{equation}
\label{Vhat_j decomp}
\frac{J}{J-1}\hat\biV_J^{\rm JK} = \sum_{j=1}^J 
   (\hat\bbeta^{(j)} - \hat\bbeta )(\hat\bbeta^{(j)} - \hat\bbeta )^\top
=N^{-2}\sum_{j=1}^J \nabla f (\bxi_j )^\top \biW_j \biW_j^\top 
\nabla f (\bxi_j ),
\end{equation}
where the factor $J/(J-1) \to 1$ will be ignored from now on.

By \Cref{assn:yap}(e), $\nabla f$ is continuous at~$\bmu$. Thus,
\eqref{lln1}, \eqref{lln5}, and Slutsky's theorem imply that 
$\nabla f(\bxi_j) = \nabla f(\bmu) +\biR_j$, where $\nabla f (\bmu ) 
= (\bzero ,\biS_{xx}^{-1})$ and $\max_{1\leq j \leq J}\Vert\biR_j\Vert 
= o_P(1)$. From \eqref{Vhat_j decomp} we then find that
\begin{align}
 \nonumber
N^2 \hat\biV_J^{\rm JK} &= \big((\bzero ,\biS_{xx}^{-1})+o_P(1)\big)^\top
\Big(\sum_{j=1}^J\biW_j\biW_j^\top\Big)\big((\bzero,
\biS_{xx}^{-1})+o_P(1)\big)\\
&=\biS_{xx}^{-1}\Big(\sum_{j=1}^J\bis_j\bis_j^\top\Big)
\biS_{xx}^{-1} +o_P \Big(\sum_{j=1}^J N_j^2 \Big),
\label{VhatJKlimit}
\end{align}
where the second line follows from \Cref{lem:orders}.

\subsubsection*{Proof of part~(i)}

\emph{Proof for~$\hat\biV_3^{(3)}$}:
Combining \eqref{lln1} and \eqref{VhatJKlimit} shows that
\begin{align}
\label{part i proof}
(\Var(\hat\bbeta))^{-1}\hat\biV_3^{(3)} 
&= (\biX^\top\biX)\bSigma^{-1}(\biX^\top\biX)
(\hat\biV_G^{\rm JK}+\hat\biV_H^{\rm JK}-\hat\biV_I^{\rm JK}) \\
\nonumber
&= \biS_{xx}\bSigma^{-1}\biS_{xx}\big(\bfI_k + o_P(1)\big) 
\biS_{xx}^{-1}\Big( \sum_{g=1}^G\bis_g\bis_g^\top 
+ \sum_{h=1}^H\bis_h\bis_h^\top -\sum_{g=1}^G\sum_{h=1}^H
   \bis_{gh}\bis_{gh}^\top \Big) \biS_{xx}^{-1} \\
\nonumber
&\quad + o_P\bigg( \lambda_N^{-1} \Big( \sum_{g=1}^G N_g^2 
   + \sum_{h=1}^H N_h^2 \Big) \bigg) \Pto \bfI_k ,
\nonumber
\end{align}
where the final convergence follows from \eqref{lln4} and 
\Cref{assn:yap}(c).

\emph{Proof for~$\hat\biV_3^{(3+)}$}: 
Under \Cref{assn:yap}, $\hat\biV_3^{(3)}$ is positive definite in the 
limit, so the result for $\hat\biV_3^{(3+)}$ is immediate.

\emph{Proof for~$\hat\biV_{3,1}^{(3)}$}:
Replacing $\hat\biV_I^{\rm JK}$ with $\hat\biV_I$ in 
\eqref{part i proof}, the next line follows unchanged by using 
\eqref{lln1}, \eqref{VhatJKlimit}, and the corresponding result for 
$\hat\biV_I$ in the proof of Proposition~2 in \citet{yap_2025}.

\subsubsection*{Proof of part~(ii)}

By minor modifications to the consistent variance estimation arguments 
in the proof of Proposition~2 in \citet{yap_2025}, it follows that
\begin{align}
\nonumber
\big(\!\Var(\bia^\top\hat\bbeta)\big)^{\!-1}
& \max \{ \bia^\top\hat\biV_1^{(3)}\bia,
\bia^\top\hat\biV_G\tk\bia , \bia^\top\hat\biV_H\tk\bia \} \\
\nonumber
&= \big(\!\Var(\bia^\top\hat\bbeta)\big)^{\!-1}
\max \Big\{ \bia^\top \biS_{xx}^{-1}
\Big( \sum_{g=1}^G \bis_g \bis_g^\top + \sum_{h=1}^H \bis_h \bis_h^\top 
+ \sum_{g=1}^G \sum_{h=1}^H \bis_{gh} \bis_{gh}^\top \Big) 
\biS_{xx}^{-1}\bia , \\
\label{max1}
&\qquad\quad \bia^\top \biS_{xx}^{-1} \Big( \sum_{g=1}^G \bis_g \bis_g^\top 
\Big) \biS_{xx}^{-1}\bia ,
\bia^\top \biS_{xx}^{-1} \Big( \sum_{h=1}^H \bis_h \bis_h^\top \Big) 
\biS_{xx}^{-1}\bia
\Big\}\big(1+o_P(1)\big).
\end{align}
By application of \eqref{VhatJKlimit},
$(\!\Var(\bia^\top\hat\bbeta))^{\!-1} \max \{
\bia^\top\hat\biV_3^{(3)}\bia, \bia^\top\hat\biV_G^{\rm JK}\bia,
\bia^\top\hat\biV_H^{\rm JK}\bia \}$ is equal to the right-hand side
of \eqref{max1} up to $o_P(1)$ terms, which proves the result.

\subsubsection*{Proof of part~(iii)}

If $\biV_I$ is asymptotically negligible in \eqref{Vtrue} then 
$\bSigma =(\sum_{g=1}^G\bSigma_g +\sum_{h=1}^H\bSigma_h)(\bfI_k +o(1))$,
and \eqref{lln4} is modified accordingly. Combining this with 
\eqref{lln1}, \eqref{VhatJKlimit}, and \Cref{assn:yap}(c) shows that
\begin{align*}
\big(\!\Var(\hat\bbeta)\big)^{\!-1}\hat\biV_3^{(2)} 
&= (\biX^\top\biX)\bSigma^{-1}(\biX^\top\biX)
(\hat\biV_G^{\rm JK}+\hat\biV_H^{\rm JK}) \\
&= \biS_{xx}\bSigma^{-1}\biS_{xx} \big(\bfI_k + o_P(1)\big) 
\biS_{xx}^{-1}\Big( \sum_{g=1}^G\bis_g\bis_g^\top 
+ \sum_{h=1}^H\bis_h\bis_h^\top \Big) \biS_{xx}^{-1} \\ 
&\quad + o_P\Big( \lambda_N^{-1} \Big( \sum_{g=1}^G N_g^2 
   + \sum_{h=1}^H N_h^2 \Big) \Big) \Pto \bfI_k .
\end{align*}

\section{Appendix B: The twowayjack Package}
\label{app:software}

\def\hangpara{\hangindent=\parindent\hangafter=1\noindent}

We have written a package called \texttt{twowayjack} for
\texttt{Stata} that implements the variance estimators discussed in
this paper. The package relies on our earlier package
\texttt{summclust} \citep{MNW-influence}, and it calculates both
CV$_{\tn3}^{(\max)}$ and CV$_{\tn1}^{(\max)}$ for the coefficient of
interest, as well as confidence intervals and $P$~values.

The package also provides coefficients of variation for several 
diagnostic measures as described in \citet{MNW-influence}. For the two
primary clustering dimensions and their intersections, it calculates
the coefficients of variation for the cluster sizes, leverage, partial
leverage, and omit-\tkk one\tkk-cluster estimates, $\hat\beta^{(g)}$, 
$\hat\beta^{(h)}$, and~$\hat\beta^{(i)}$. In addition, it displays
the number of clusters $G$ and the effective number of clusters
$G^*=G^*(0)$ from \citet{CSS_2017}.  The latest version may be obtained
from \url{https://github.com/mattdwebb/twowayjack}. The data and
programs used in the paper may be found at
\url{http://qed.econ.queensu.ca/pub/faculty/mackinnon/twowayjack/}.

\subsection{Syntax}

The syntax for \texttt{twowayjack} is

\smallskip

\texttt{
\hangpara twowayjack varlist, \underbar{clus}ter(varlist)
[\underbar{fevar}(varlist) \underbar{sam}ple(string)]
}

\noindent Here \texttt{varlist} contains a list of variables. The
first one is the dependent variable, the second is the regressor for
which standard errors and $P$~values are to be calculated, and the
remaining ones are all the other continuous and binary regressors.
Categorical variables to be treated as fixed effects should be listed 
using the \texttt{\underbar{fevar}} option.

\smallskip

\texttt{\underbar{clus}ter(varlist)} is mandatory, where
\texttt{varlist} contains the two variables by which observations are
clustered. The program returns an error if exactly two variables
are not specified.

\smallskip

\texttt{\underbar{fevar}(varlist)}. Categorical variables to be
included in the model as fixed effects should be listed here. They are
handled equivalently to \texttt{i.varlist} in a regression model.
Since this option uses a generalized inverse, CV$_{\tn3}$ can be
calculated even when some of the omit-\tkk one\tkk-cluster subsamples
are singular. This always happens with cluster-level fixed effects. 

\smallskip

\texttt{\underbar{sam}ple(string)} limits the sample. Use the text you 
would enter after an ``if'' in a regression command. For instance,
\texttt{sample(female==1)} is equivalent to ``\texttt{if female==1}.''

\subsection{Illustration}

In this section, we demonstrate how to use the package with an example
from the \texttt{webuse} dataset \texttt{nlswork}. The outcome of
interest is hours worked. The independent variable of interest is
\texttt{vismin}, which is set to 0 if a person is white, and 1
otherwise.  In part because hours vary with age, and across industry,
we cluster by both age and industry. 

The first commands load and clean the dataset.

\begin{verbatim}	
	webuse nlswork, clear
	keep if inrange(age,25,35)
	gen vismin = inrange(race,2,3)
\end{verbatim}

\noindent For comparison purposes, the native \texttt{Stata} regression 
results with one\tkk-way clustering are obtained from the command:

\medskip

\noindent
\texttt{reg hours vismin south i.age i.birth\_yr i.year i.ind , cluster(ind)
}

\medskip

\noindent This yields the results:

\begin{verbatim}
Linear regression                               Number of obs     =     13,754
                                                F(10, 11)         =          .
                                                Prob > F          =          .
                                                R-squared         =     0.0659
                                                Root MSE          =     9.6979
                              (Std. err. adjusted for 12 clusters in ind_code)
------------------------------------------------------------------------------
             |               Robust
       hours | Coefficient  std. err.      t    P>|t|     [95% conf. interval]
-------------+----------------------------------------------------------------
      vismin |   1.054672   .4202197     2.51   0.029     .1297746    1.979569
[additional output truncated]
\end{verbatim}

\medskip

We also estimate the model with two-way clustering using the native 
\texttt{Stata} command:

\medskip

\noindent
\texttt{reg hours vismin south i.age i.birth\_yr i.year i.ind , vce(cluster age ind) 
}

\medskip

\noindent This yields the results:

\begin{verbatim}
note: multiway-cluster variance–covariance matrix is not positive semidefinite.

Linear regression                                     Number of obs = 13,754
Clusters per comb.:                                   Cluster comb. =      3
min =  11                                             F(18, 10)     =      .
avg =  52                                             Prob > F      =      .
max = 132                                             R-squared     = 0.0659
Adj R-squared = 0.0625
Root MSE      = 9.6979

[table continues on next page]
\end{verbatim}

\begin{verbatim}

(Std. err. adjusted for multiway clustering)
------------------------------------------------------------------------------
       |               Robust
hours  | Coefficient  std. err.      t    P>|t|     [95% conf. interval]
-------+----------------------------------------------------------------
vismin |   1.054672   .4372782     2.41   0.037     .0803553    2.028988
[additional output truncated]
\end{verbatim}

Note the warning message about the non-positive\tkk-semidefinite error 
variance matrix. To highlight the issue of the eigenvalue correction, 
we re-estimate the same model, but hard-code the industry fixed effects
to omit the dummy corresponding to the most common industry.

\begin{verbatim}
 	tab ind, gen(ins)
 	reg hours vismin south i.age i.birth_yr i.year \\\
 	ins1-ins10 ins12 , vce(cluster age ind)
\end{verbatim}

\medskip

\noindent This yields the truncated results:

\begin{verbatim}
------------------------------------------------------------------------------
       |               Robust
 hours | Coefficient  std. err.      t    P>|t|     [95% conf. interval]
-------+----------------------------------------------------------------
vismin |   1.054672   .4320889     2.44   0.035     .0919177    2.017426
[additional output truncated]
\end{verbatim}

\noindent Notice that, although the coefficient is unchanged, the
standard error, and the corresponding $P$~value and confidence
interval, are different. While inferences remain unchanged here, it is
easy to see how this type of difference could change the results of
hypothesis tests at conventional levels of significance.

We can instead estimate the same model with two-way clustering using
\texttt{twowayjack}:

\medskip

\noindent
\texttt{twowayjack hours vismin south , fevar(age birth\_yr year ind)
cluster(age ind)}

\medskip

\noindent The output is:

\begin{verbatim}
TWOWAYJACK
Reference: James G. MacKinnon, Morten Ø. Nielsen, and Matthew D. Webb
           Jackknife inference with two-way clustering
	   
Two-way cluster jackknife variance estimation.
Cluster summary statistics for vismin when clustered by age and ind_code.

Regression Output
    s.e. |      Coeff   Sd. Err.   t-stat  P value    CI-lower    CI-upper
---------+----------------------------------------------------------------
  CV1max |   1.054672   0.420220   2.5098   0.0309    0.223377    1.885967
  CV3max |   1.054672   0.521628   2.0219   0.0708   -0.107587    2.216931
--------------------------------------------------------------------------
\end{verbatim}
\goodbreak
\begin{verbatim}
Coefficients of Variation, G, and G*
 dimension |       Ng   Leverage  Partial L.   beta no g        G      Gstar
-----------+----------------------------------------------------------------
       age |   0.0987     0.1813      0.0927      0.0431       11      10.90
  ind_code |   1.1815     0.8823      1.1849      0.1565       12       5.21
 intersect |   1.1507     0.8925      1.1557      0.0173      132      56.26
----------------------------------------------------------------------------
\end{verbatim}

\noindent It may seem odd that the CV$_{\tn1}^{\rm max}$ standard error
reported here (0.420220) is smaller than the native \texttt{Stata} 
two\tkk-way standard error reported earlier~(0.437278). This happens 
because the latter is actually the CV$_{\tn1}^{3+}$ standard error, 
and the variance matrix is evidently not positive definite in this case.

\setlength{\bibsep}{0pt}
\bibliography{mnw-twoway}
\addcontentsline{toc}{section}{\refname}

\end{document}